\documentclass{lmcs}
\pdfoutput=1

\usepackage{lastpage}
\lmcsdoi{21}{4}{14}
\lmcsheading{}{\pageref{LastPage}}{}{}%
{Oct.~29,~2024}{Oct.~28,~2025}{}


\usepackage[utf8]{inputenc}
\usepackage[english]{babel}
\usepackage{amsmath,amssymb}
\usepackage[hyphens]{url}
\usepackage{hyperref}
\usepackage{mathpartir}

\newcommand{\weg}[1]{}

\newcommand{\set}[1]{\left\{ #1 \right\}}
\newcommand{\compr}[2]{\left\{ #1 \mid #2 \right\}}
\newcommand{\tuple}[1]{\langle #1 \rangle}
\renewcommand{\phi}{\varphi}

\newcommand{\bis}{\mathrel{\underline{\leftrightarrow}}}
\newcommand{\dbis}{\mathrel{\underline{\rightarrow}}}
\newcommand{\apt}{\mathrel{\underline{\#}}}
\newcommand{\dapt}{\mathrel{\underline{\#\!\!\!{>}}}}
\newcommand{\toop}{\mathbin{\to}}

\newcommand{\bBisim}{\mathrel{\bis_b}}
\newcommand{\sBisim}{\mathrel{\bis_s}}
\newcommand{\dsbBisim}{\mathrel{\dbis_{dsb}}}
\newcommand{\dsBisim}{\mathrel{\dbis_{ds}}}

\newcommand{\dBisim}{\mathrel{\dbis_{db}}}
\newcommand{\dBisimSym}{\mathrel{\bis_{db}}}

\newcommand{\dsApart}{\mathrel{\dapt_{ds}}}

\newcommand{\dApart}{\mathrel{\dapt_{db}}}
\newcommand{\dApartSym}{\mathrel{\apt_{db}}}

\newcommand{\maybeto}[1]{\to_{(#1)}}
\newcommand{\toalpha}{\to_\alpha}
\newcommand{\maybetoalpha}{\maybeto{\alpha}}
\newcommand{\totau}{\to_\tau}

\newcommand{\rtc}{\twoheadrightarrow}
\newcommand{\rtctau}{\rtc_\tau}

\newcommand{\eventual}[4]{#2 \rtctau #3 \to_{(#1)} #4}
\newcommand{\eventualalpha}[3]{\eventual{\alpha}{#1}{#2}{#3}}
\newcommand{\commoneventual}[1]{\eventualalpha{#1}{#1'}{#1''}}
\newcommand{\sateventual}[5]{#2 \rtctau^{#5} #3 \to_{(#1)} #4}
\newcommand{\satcommoneventual}[2]{\sateventual{\alpha}{#1}{#1'}{#1''}{#2}}

\newcommand{\Th}{{\normalfont\textrm{Th}}}
\newcommand{\ThP}{{\normalfont\textrm{Th}_{\normalfont\textrm{P}}}}
\newcommand{\HML}{{\normalfont\textrm{HML}}}
\newcommand{\PHML}{{\normalfont\textrm{PHML}}}
\newcommand{\pmPHML}{\pm\PHML}
\newcommand{\cpmPHML}{\bigwedge\pmPHML}
\newcommand{\HMLU}{{\normalfont\textrm{HMLU}}}
\newcommand{\PHMLU}{{\normalfont\textrm{PHMLU}}}
\newcommand{\pmPHMLU}{\pm\PHMLU}
\newcommand{\cpmPHMLU}{\bigwedge\pmPHMLU}

\newcommand{\sizeChoice}{}

\DeclareMathOperator{\kwpos}{pos}
\DeclareMathOperator{\kwneg}{neg}
\DeclareMathOperator{\NRS}{NRS}

\DeclareMathOperator{\LTS}{LTS}
\DeclareMathOperator{\LTStau}{LTS_\tau}

\begin{document}
  \title{Positive Hennessy-Milner Logic for Branching Bisimulation}

  \author[H.~Geuvers]{Herman Geuvers\lmcsorcid{0000-0003-2522-2980}}[a]
  \author[K.~Golov]{Komi Golov\lmcsorcid{0000-0001-9273-413X}}[b]

  \address{Radboud University Nijmegen, The Netherlands}
  \email{herman@cs.ru.nl}
  \address{JetBrains, The Netherlands}
  \email{komi.golov@jetbrains.com}

  \begin{abstract}
    Labelled transitions systems can be studied in terms of modal logic
    and in terms of bisimulation.
    These two notions are connected by Hennessy-Milner theorems,
    that show that two states are bisimilar precisely when they satisfy the same modal logic formulas.
    Recently, {\em apartness\/} has been studied as a dual to
    bisimulation, which also gives rise to a dual version of the
    Hennessy-Milner theorem: two states are apart precisely when there
    is a modal formula that distinguishes them.

    In this paper, we introduce ``directed'' versions of
    Hennessy-Milner theorems that characterize when the theory of one
    state is included in the other.
    For this we introduce ``positive modal logics'' that only allow a limited
    use of negation. Furthermore, we introduce directed notions of
    bisimulation and apartness, and then show that, for this positive
    modal logic, the theory of $s$ is included in the theory of $t$
    precisely when $s$ is directed bisimilar to $t$. Or, in terms of
    apartness, we show that $s$ is directed apart from $t$
    precisely when the theory of $s$ is not included in the theory of
    $t$. From the directed version of the Hennessy-Milner theorem, the
    original result follows.

    In particular, we study the case of branching
    bisimulation and Hennessy-Milner Logic with Until (HMLU) as a
    modal logic. We introduce ``directed branching bisimulation'' (and
    directed branching apartness) and ``Positive Hennessy-Milner Logic
    with Until'' (PHMLU) and we show the directed version of the
    Hennessy-Milner theorems.  In the process, we show that every HMLU
    formula is equivalent to a Boolean combination of Positive HMLU formulas, which is a very
    non-trivial result.  This gives rise to a sublogic of HMLU that is
    equally expressive but easier to reason about.
  \end{abstract}

  \maketitle
\noindent 
  Concurrent systems are usually modeled as labelled transition
  systems (LTS), where a labelled transition step takes one from one
  state to the other in a non-deterministic way. Two states in such a
  system are considered equivalent in case the same labelled
  transitions can be taken to equivalent states. This is captured via
  the notion of bisimulation, which is a type of observable equality:
  two states are bisimilar if we can make the same observations on
  them, where the observations are the labelled transitions to
  equivalent states.

  Often it is the case that some transitions cannot be observed and
  then they are modeled via a silent step, also called a $\tau$-step,
  denoted by $t \to_{\tau} s$. The notion of bisimulation should take
  this into account and abstract away from the silent steps. This
  leads to a notion of weak bisimulation (as opposed to strong
  bisimulation, where all steps are observable). In \cite{GlabbeekW89}
  (and the later full version \cite{GlabbeekWeijland}) it has been
  noticed that weak bisimulation ignores the branching in an LTS that
  can be caused by silent steps, and that is undesirable. Therefore
  the notion of branching bisimulation has been developed, which
  ignores the silent steps while taking into account their branching
  behavior. Branching bisimulation has been studied a lot as it is
  the finest equivalence in the well-known ``van Glabbeek spectrum''
  \cite{GlabbeekSpectrum}. Various algorithms have been developed for
  checking branching bisimulation and various tools have been
  developed that verify branching bisimulation for large systems
  \cite{GrooteVaandrager,JansenEtal}.

  There is a well-known connection between notions of bisimulation and
  various modal logics, first discovered by Hennessy and Milner~\cite{HennessyMilner}
  for strong bisimulation, and then extended to weak and branching
  bisimulation by De Nicola and Vaandrager~\cite{NicolaVaandrager}.
  Let $\bis$ be a bisimilarity relation, and let $\Th(s)$ be the set of
  formulas in some modal logic that are true in a state $s$.
  A Hennessy-Milner theorem is a theorem relating the bisimulation and
  the logic by showing that two states are bisimilar precisely when their
  theories coincide:
  \[
    s \bis t \iff \Th(s) = \Th(t).
  \]

  Hennessy and Milner proved such a theorem for Hennessy-Milner Logic and
  strong bisimulation, while De Nicola and Vaandrager extended this to
  Hennessy-Milner Logic\footnote{This involved modifying the semantics
  to account for silent steps.} and weak bisimulation, and
  Hennessy-Milner Logic with Until and branching bisimulation.

  In this paper, we explore what a directed variant of such a theorem
  would look like by considering what form of ``directed
  bisimulation'' relation corresponds to inclusion of theories.
  Hennessy-Milner logics typically have negation, so if the theory of
  one state is included in the other, then they are equal. As a
  consequence, Hennessy-Milner logics typically do not permit
  non-trivial inclusions of theories, so we must first replace the
  logic with a positive version.  Writing $\dbis$ for our new relation
  and $\ThP(s)$ for the theory of $s$ in this new, positive, logic, we
  prove the property
  \begin{equation}
    \label{eqn:directedHM}
    s \dbis t \iff \ThP(s) \subseteq \ThP(t).
  \end{equation}

  Our constructions ensure that $s \bis t$ precisely when $s \dbis t$ and
  $t \dbis s$, and that $\ThP(s) = \ThP(t)$ precisely when $\Th(s) = \Th(t)$.
  The undirected Hennessy-Milner theorem is thus an immediate consequence of our
  directed variant.

  Our proof of Equation~(\ref{eqn:directedHM}) uses the notion of apartness,
  which is the dual notion of bisimulation,
  studied previously by Geuvers and Jacobs~\cite{GeuversJacobs}.
  Two states are bisimilar precisely when they are not apart.
  The property we prove is thus
  \begin{equation}
    \label{eqn:directedHM-apartness}
    s \dapt t \iff \exists \varphi.\, \varphi \in \ThP(s) - \ThP(t),
  \end{equation}
  where $\varphi$ is called the distinguishing formula.

  Unlike bisimulation, apartness is an inductive notion, meaning that we can
  perform our proofs by an induction on the derivation of an apartness.
  It turns out that the structure of this derivation closely mirrors the distinguishing
  formula, making the proof straightforward.
  This was previously shown by Geuvers in relation to strong and weak bisimulation
  in~\cite{Geuvers}, but a straightforward inductive proof does not work in the
  branching setting when the logic includes an \emph{until} operator.

  In this paper we study the directed constructions for branching bisimulation.
  Concretely, we define and study (Section~\ref{sec.HMLU})
  a logic of Positive Hennessy-Milner Logic with Until formulas
  which we refer to as PHMLU (Subsection~\ref{ssec.PHMLU}),
  as well as a directed branching bisimulation relation $\dBisim$ (Subsection~\ref{ssec.dirbranbis})
  and a directed branching apartness relation $\dApart$ (Subsection~\ref{ssec.dirbranapt}).
  We then show that Equation~(\ref{eqn:directedHM-apartness})
  (for PHMLU and branching apartness) holds,
  and from that we conclude that Equation~(\ref{eqn:directedHM}) holds as well.
  There are differences between HMLU and PHMLU that make it far from obvious
  that the logics are equally powerful.
  We show that the logics are equivalent in which states they distinguish,
  and moreover we show that every HMLU formula is equivalent to a Boolean
  combination of PHMLU formulas (Subsection~\ref{ssec.PHMLU}).
  This gives rise to a logic that is equally expressive as HMLU, but for which
  the satisfaction relation is simpler.

  To introduce the directed versions of bisimulation and apartness,
  and the corresponding positive modal logics, we will first start
  from the simpler setting of strong bisimulation and standard
  Hennessy-Milner Logic in Section~\ref{sec.strong}. Here the
  construction and the proofs are rather straightforward, but it also
  suggests that a `directed approach' to bisimulation, combined with a
  positive version of Hennessy-Milner logic and the use of apartness,
  should be applicable to other notions of bisimulation. For weak
  bisimulation, this is the case, which we have verified, but not
  included in the present paper.
      
  Throughout this paper, we will work in a fixed labelled transition system (LTS)
  $(X, A, \to)$, consisting of a set of \emph{states} $X$, a set of
  \emph{actions} $A$, possibly with a \emph{silent action} $\tau \not\in A$,
  and a transition relation $\to_a$ for every $a \in A$ (and for $\tau$, if applicable).
  Following~\cite{GeuversJacobs}, we will use $\alpha$ to denote an arbitrary
  action from $A \cup \set{\tau}$, $a$ to denote an arbitrary non-silent action (so $a\in A$), and
  use $\rtc_\alpha$ to denote the transitive reflexive closure of $\toalpha$.
  We use $\maybetoalpha$ as a shorthand for $\to_a$ if $\alpha = a \neq \tau$,
  and for the reflexive closure of $\to_\tau$ if $\alpha = \tau$.
  
  We would like to thank Jan Friso Groote and Jurriaan Rot for
  discussions on the topic of this paper. We also thank the referees, in particular also the referees of earlier versions of this paper, for their useful comments. 

  \section{Hennessy-Milner Logic and Strong Bisimulation}
\label{sec.strong}
  \begin{defi}
    A \emph{labelled transition system} (LTS) is a tuple $(X, A, \toop)$, where $X$
    is a set of \emph{states}, $A$ is a set of \emph{actions} and for every $a \in A$, $\mathbin{\to_a}$ is a binary relation on $X$.
    When $s \to_a t$ holds, we say that $s$ \emph{$a$-steps to} $t$.
    
    An LTS is \emph{image-finite} if for all $s \in X$ and $a \in A$,
    the set $\compr{t \in X}{s \to_a t}$ is finite.
  \end{defi}

  We will be particularly interested in image-finite LTSs, since the
  logics we study are finitary. 
  We can characterize states in an LTS using Hennessy-Milner Logic (HML).

  \begin{defi}
    The formulas of HML are given inductively by the following definition:
    \[
      \varphi := \top \mid \neg \varphi \mid \varphi_1 \wedge \varphi_2 \mid \tuple{a} \varphi.
    \]
  \end{defi}

  We use $\bot$ as a shorthand for $\neg\top$ and
  $\varphi_1 \vee \varphi_2$ as a shorthand for $\neg(\neg\varphi_1 \wedge \neg\varphi_2)$.
  The modality binds weakly; $\tuple{a} \varphi \wedge \psi$ denotes $\tuple{a} (\varphi \wedge \psi)$.

  \begin{defi}
    \label{def.strong_interp}
    The relation $s \vDash \varphi$ is defined inductively by the following rules:
    $$\begin{array}{lclclcl}
      s \vDash \top &&\text{always} &\qquad\qquad\qquad&
      s \vDash \varphi_1 \wedge \varphi_2 &\Leftrightarrow &\text{$s \vDash \varphi_1$ and $s \vDash \varphi_2$}\\
      s \vDash \neg \varphi &\Leftrightarrow &s \not\vDash \varphi & &
            s \vDash \tuple{a} \varphi &\Leftrightarrow
          &\exists s'.\, s \to_a s' \wedge s' \vDash \varphi.
    \end{array}$$
\noindent 
    Given a state $s$ we define $\Th(s) := \compr{\varphi \in \HML}{s \vDash \varphi}$, the set of HML formulas 
    true in $s$.
    We call a formula in $\Th(s) - \Th(t)$ or $\Th(t) - \Th(s)$ a \emph{distinguishing formula} for $s$ and $t$.
  \end{defi}

  To compare two states, one 
  considers if they can recursively simulate one another.
  This notion is known as bisimulation.

  \begin{defi}
    A symmetric relation $R$ is a strong bisimulation if whenever $R(s,t)$ and $s \to_a s'$,
    there exists some $t'$ such that $R(s',t')$ and $t \to_a t'$.

    We say two states are strongly bisimilar (notation $s \sBisim t$) if
    there exists a strong bisimulation that relates them.
  \end{defi}

  A key result of Hennessy and Milner~\cite{HennessyMilner} is that these two views of
  labelled transition systems are related in the following sense:

  \begin{thmC}[\cite{HennessyMilner}]
    In an image-finite LTS, two states $s$ and $t$ are strongly bisimilar precisely when they
    satisfy the same HML formulas; that is,
    \[
      s \sBisim t \Leftrightarrow \Th(s) = \Th(t).
    \]
  \end{thmC}
\noindent 
  We are interested in a similar theorem that considers inclusion of theories rather than equality.
  However, in the context of Hennessy-Milner Logic such a theorem would not be meaningfully different,
  since the presence of negation gives us the following result:

  \begin{prop}
    If $\Th(s) \subseteq \Th(t)$, then $\Th(s) = \Th(t)$.
  \end{prop}

  \begin{proof}
    Suppose, on the contrary, that there is some $\varphi \in \Th(t) - \Th(s)$. 
    Then $\neg\varphi \in \Th(s) \subseteq \Th(t)$, hence $t \vDash \varphi$ and $t \vDash \neg\varphi$,
    a contradiction.
  \end{proof}

  We will thus introduce Positive Hennessy-Milner Logic (PHML),
  where negation is only permitted under a modality.\footnote{
  Prohibiting negation entirely would produce a logic that is too weak;
  it would correspond to simulation.
  Two states simulating each other is weaker than two states being bisimilar,
  while we require that two states directed bisimulating each other be
  equivalent to the states being bisimilar.}

  \begin{defi}
    The formulas of PHML are given inductively by the following definition:
    \[
      \varphi := \top \mid \bot \mid \varphi_1 \wedge \varphi_2 \mid \varphi_1 \vee \varphi_2 \mid \tuple{a} (\varphi_1 \wedge \neg \varphi_2).
    \]
  \end{defi}

  We view a formula of PHML as also being a formula of HML, and can thus
  reuse the semantics from Definition~\ref{def.strong_interp}.
  We define $\ThP(s) := \compr{\varphi \in \PHML}{s \vDash \varphi}$.
  We call a $\phi \in \ThP(s) - \ThP(t)$ a \emph{positive distinguishing formula} for $s$ and $t$.
  
  There are various syntactic classes related to PHML that will be
  useful for our proofs, and for these we introduce some special
  notation.
  
  \begin{nota}\label{nota.phml}\hfill
    \begin{itemize}
    \item $\pmPHML$ denotes the set of PHML formulas and their negations.
    \item $\cpmPHML$ denotes the set of conjunctions of $\pmPHML$ formulas.
    \item Given a formula $\varphi \in \cpmPHML$, we will use
      $\varphi^+$ and $\varphi^-$ to refer to the PHML formulas such
      that $\varphi$ is equivalent to $\varphi^+ \wedge
      \neg\varphi^-$.
    \end{itemize}
  So $\varphi^+$ is the conjunction of all `positive' conjuncts of $\varphi$, the {\em positive segment\/} of $\phi$,
  and $\varphi^-$ is the disjunction of $\psi$ for every `negative' conjunct $\neg\psi$ of $\varphi$, the {\em negative segment\/} of $\phi$. Concretely, for $\varphi \in \cpmPHML$ with $\varphi = \varphi_1 \wedge  \varphi_2 \wedge \neg\varphi_3 \wedge  \neg\varphi_4$,  we have $\varphi^+ = \varphi_1 \wedge  \varphi_2$ and $\varphi^- =\varphi_3 \vee  \varphi_4$, and so $\varphi$ is equivalent to $\varphi^+ \wedge \neg \varphi^-$.
  We write $\tuple{a}\varphi$ for $\tuple{a}\varphi^+ \wedge \neg \varphi^-$.
  Recall that by our conventions, this is read as $\tuple{a}(\varphi^+ \wedge \neg \varphi^-)$.
  \end{nota}
  
  \begin{exa}\label{exa.nontrivial} If we consider PHML, we can have non-trivial inclusion of theories. This is illustrated by the following LTS.
    (The fact the inclusions hold in the first place is most easily shown by constructing a strong directed
    bisimulation as defined below.  However, the rigorous reader can perform a proof by induction.)

  {\sizeChoice
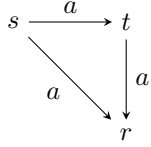
    \begin{center}
      \begin{tikzpicture}[>=stealth,node distance=1.5cm,auto]
          \node  (s)                         {$s$};
          \node  (t) [right of = s]          {$t$};
          \node  (r) [below of = t]          {$r$};
          \path[->]
            (s)   edge    node {$a$} (t)
                  edge    node[swap] {$a$}  (r)
            (t)  edge    node {$a$}  (r);
      \end{tikzpicture}
    \end{center}
  }

    In this LTS, we have $\ThP(r) \subsetneq \ThP(t) \subsetneq \ThP(s)$.
    Examples of PHML formulas that distinguish $s$, $t$ and
    $r$ are the following.
  \begin{itemize}
  \item $\tuple{a}\tuple{a}\top$, which holds in $s$ but not in $t$ and $r$.
  \item $\tuple{a}\neg\tuple{a}\top$ and $\tuple{a}\top$, which hold in $s$ and $t$, but not in $r$
  \end{itemize}
Note that $\tuple{a}\bot$ does not hold in any state, but for different reasons: in $r$ there is no $a$-step possible, whereas in $s$ and $t$ there is an $a$-step possible, but not to a state where $\bot$ holds (because $\bot$ never holds). 
  \end{exa}

  \begin{thm}
    \label{thm.strong_theory_equiv}
    Every HML formula $\varphi$ is equivalent to a disjunction of $\cpmPHML$ formulas.
  \end{thm}

  \begin{proof}
    The proof proceeds by induction on $\varphi$. We treat the cases for $\varphi$ being a negation and for $\varphi$ being a modality.
    
    Suppose that $\varphi = \neg \psi$, then by induction there is a
    formula $\bigvee_{i=1}^n \Psi_i$ equivalent to $\psi$ and we write
    $\Psi_i = \Psi_i^+ \wedge \neg \Psi_i^-$ (for $i\in
    \{1,\ldots,n\}$). Then $\varphi = \neg \psi$ is equivalent to
    $\bigwedge_{i=1}^n \neg(\Psi_i^+) \vee \Psi_i^-$. Note that
    $\Psi_i^- \in \PHML$, but $\neg(\Psi_i^+)$ is a negation of a
    $\PHML$-formula. We can let $\wedge$ distribute over $\vee$ and
    we find that $\varphi$ is equivalent to
    a disjunction of $\cpmPHML$ formulas.
    
    Suppose that $\varphi = \tuple{a} \psi$, then by induction there is a
    formula $\bigvee_{i \in I} \Psi_i$ equivalent to $\psi$.
    Since modality distributes over disjunction, $\tuple{a} \bigvee_i \Psi_i$
    is equivalent to $\bigvee_i \tuple{a} \Psi_i$, which is the desired formula.
  \end{proof}

  \begin{cor}
    \label{cor:strong_theory_equality}
    For all states $s, t$, we have $\Th(s) = \Th(t)$ precisely when $\ThP(s) = \ThP(t)$.
  \end{cor}

  \begin{proof}
    The left to right direction is easy: $\ThP(s)= \Th(s) \cap \PHML$,
    so $\Th(s) = \Th(t)$ implies $\ThP(s)= \ThP(t)$.

    In the other direction, suppose $\ThP(s) = \ThP(t)$ and $\varphi \in \Th(s)$.
    By Theorem~\ref{thm.strong_theory_equiv}, there is an equivalent formula
    $\bigvee_i \Psi_i$ where $\Psi_i \in \cpmPHML$. As $s\vDash \phi$,
    there is some $i$ such that $s \vDash \Psi_i$.
    Writing $\Psi_i = \bigwedge_j \Psi^+_j \wedge \neg \Psi^-_j$, we have that for all $j$,
    $s \vDash \Psi^+_j$ and $s \not \vDash \Psi^-_j$.
    Since each $\Psi^+_j$ and $\Psi^-_j$ is positive, it follows that
    $t \vDash \Psi^+_j$ and $t \not \vDash \Psi^-_j$
    and thus $t \vDash \Psi_i$ and $t \vDash \varphi$, as required.
  \end{proof}

  This concludes our construction on the logic side.
  Let us now consider the bisimulation side of the question.

  \begin{defi}
    A relation $R$ is a \emph{directed strong bisimulation} if whenever $R(s, t)$ and $s \to_a s'$,
    there exists some $t'$ such that $t \to_a t'$ and $R(s',t')$ and $R(t',s')$. 

    We say two states are \emph{directed strongly bisimilar} (notation $s \dsBisim t$) if
    there exists a directed strong bisimulation that relates them.
  \end{defi}

  \begin{thm}
    Two states are strongly bisimilar if and only if they are directed strongly
    bisimilar in each direction.
  \end{thm}

  \begin{proof}
    The left-to-right direction follows from the fact that every strong bisimulation
    is a directed strong bisimulation.
    For the other direction, define $R(p, q) = p \dsBisim q \wedge q \dsBisim p$.
    This is a strong bisimulation, and hence $R(p, q)$ implies that $p$ and $q$
    are strongly bisimilar.
  \end{proof}

  We can now state our directed Hennessy-Milner theorem for strong bisimulation.

  \begin{thm}
    \label{thm.directed_strong_hm}
    In an image-finite LTS, for all $s, t$, $s \dsBisim t$ iff $\ThP(s) \subseteq \ThP(t)$.
  \end{thm}
\noindent 
  In order to prove this theorem we introduce one more directed notion:
  directed strong apartness, the dual of directed strong bisimulation.
  This will simplify the proof to an inductive argument in each direction, one
  induction on the formula that distinguishes the states, and the other induction on
  the derivation of the apartness.

  \begin{defi}\label{def.dirapt}
    A relation $Q$ is a \emph{directed strong apartness} if the following rule holds for $Q$.
    \[
      \inferrule*[Right=$\textsc{in}_{ds}$]{
        s \to_a s' \\
        \forall t'.\, t \to_a t' \implies Q(s', t') \vee Q(t', s')
      }{Q(s, t)}
    \]
    Two states $s, t$ are \emph{directed strong apart}, notation $s
    \dsApart t$, if for every directed strong apartness $Q$, it
    holds that $Q(s, t)$.
  \end{defi}

We want to stress that directed strong apartness is an {\em inductive
  notion}, like other notions of apartness, see~\cite{GeuversJacobs}. This implies that it can be characterized via a
     {\em derivation system}: $s \dsApart t$ if and only if this can be
     derived with the derivation rule $\textsc{in}_{ds}$.  It is also
     immediate that directed strong apartness is the complement of
     directed strong bisimulation. We collect these facts in the
     following easy to prove lemma.

  \begin{lem} \label{lem.dirsapt}
We have that $s \dsApart t$ if and only if this can be derived using the following rule.
  \[
    \inferrule*[Right=$\textsc{in}_{ds}$]{
      s \to_a s' \\
      \forall t'.\, t \to_a t' \implies s' \dsApart t' \vee t' \dsApart s'
    }{s \dsApart t}
  \]
Furthermore, $s \dsApart t \quad\Leftrightarrow \quad \neg (s\dsBisim t)$.
  \end{lem}

  \begin{proof}
    Phrased differently:  if we redefine $s \dsApart t$ as ``there is a derivation of $s \dsApart t$ using only rule $\textsc{in}_{ds}$'', then we can show that $s \dsApart t$ if and only if  $Q(s,t)$ holds for all apartness relations $Q$.
    
    The first observation is that the redefined $s \dsApart t$ is itself an apartness relation.  
    Hence, if $Q(s,t)$ for all apartness relations $Q$, then $s \dsApart t$. The other way
    around, if $s \dsApart t$ by a derivation, and $Q$ is an apartness
    relation, then we can prove $Q(s,t)$ by induction on the
    derivation.

    For the second part: For $\Rightarrow$ we derive a contradiction
    from $s \dsApart t$ and $s\dsBisim t$ by induction on the
    derivation of $s \dsApart t$. For $\Leftarrow$, we show that
    $\neg(s \dsApart t)$ is a directed bisimulation relation, so
    $\neg(s \dsApart t)$ implies $s\dsBisim t$.
  \end{proof}
  
One may wonder what the ``base case'' is for the inductive definition
of $s \dsApart t$, because every application of rule
$\textsc{in}_{ds}$ seems to have an apartness hypothesis again. In
case $s \to_a s'$ and there is no $a$-step from $t$, then the $\forall
t'.\, t \to_a t' \implies \ldots$ trivially holds, and we find that $s
\dsApart t$. \vspace{0.5\baselineskip} 

  Every two states are either directed strong bisimilar or directed strong apart,
  depending on whether there is any directed strong bisimulation that relates them.
  We can thus restate Theorem~\ref{thm.directed_strong_hm} as follows.

  \begin{thm}\label{thm.directed_strong_apt_hm}
    In an image-finite LTS, for all $s, t$, $s \dsApart t$ iff there is some
    positive distinguishing formula for $s$ and $t$.
  \end{thm}

  \begin{proof}
    Both directions of the proof proceed by induction:
    from left to right by induction on the derivation of apartness,
    and from right to left by induction on the positive distinguishing formula.

    For the left to right case, $s \dsApart t$ has been concluded by
    the $\textsc{in}_{ds}$ rule (see Lemma~\ref{lem.dirsapt}), so say
    we have $s \to_a s'$ and we know $s' \dsApart t' \vee t' \dsApart
    s'$ for all (finitely many!) $t'$ for which $t \to_a t'$. By
    induction we have, for each of these $t'$, either a formula
    $\varphi' \in \ThP(s') - \ThP(t')$ or a formula
    $\psi' \in \ThP(t') - \ThP(s')$. Taking $\Phi$ to be
    the conjunction of all $\varphi'$ and $\Psi$ to be the disjunction of
    all $\psi'$, we find that $\varphi:= \tuple{a} (\Phi \wedge \neg
    \Psi)$ is in $\ThP(s) -\ThP(t)$.

    For the right to left case, we do induction on $\phi \in
    \ThP(s)-\ThP(t)$. The interesting case is when $\varphi$ is of the form
    $\tuple{a} (\varphi_1 \wedge \neg \varphi_2)$.
    Then $s \to_a s'$ with $s' \vDash \varphi_1 \wedge \neg
    \varphi_2$ for some $s'$. Also, there is no $t'$ with $t \to_a t'$
    and $t' \vDash \varphi_1 \wedge \neg \varphi_2$. So for all $t'$
    with $t \to_a t'$ we either have $t'\not\vDash \phi_1$ or
    $t'\vDash \phi_2$. By induction we find that $\forall t'.\, t
    \to_a t' \implies s' \dsApart t' \vee t' \dsApart s'$, and we can
    conclude $s\dsApart t$ by the rule $\textsc{in}_{ds}$.
  \end{proof}
  
  \section{Hennessy-Milner Logic with Until and Branching Bisimulation}
  \label{sec.HMLU}

  Let us now extend our notion of a labelled transition system to permit silent steps.
  We extend our system with a special `silent' action $\tau \not\in A$ with a
  corresponding silent step relation $\mathbin{\totau}$.

  \begin{defi}
    A \emph{labelled transition with silent steps} ($\LTStau$) is
    a labelled transition system $(X, A, \toop)$ with a special action $\tau \not\in A$
    and a corresponding relation $\mathbin{\totau}$ on $X$.
  \end{defi}

  Our presentation of branching bisimulation follows the style of
  Basten~\cite{Basten} and van Glabbeek et
  al.~\cite{GlabbeekLuttikTricka}, which has also been described
  already in~\cite{GlabbeekSpectrum}.  We use the abbreviation $s
  \maybetoalpha t$ as a shorthand for $s \to_\alpha t \vee (s = t
  \wedge \alpha = \tau)$.
  
  We will refer to a sequence $\commoneventual{s}$ as an \emph{eventual $\alpha$-step},
  where the  $\maybetoalpha$ indicates that for $\alpha=\tau$ the last step is optional.

  We adapt our notion of image-finiteness to this new setting as follows,
  effectively requiring that for every $\alpha \in A \cup \set{\tau}$, every state
  only has finitely many outgoing eventual $\alpha$-steps.

  \begin{defi}
    An $\LTStau$ $(X, A, \toop)$ is \emph{image-finite} if the underlying LTS is
    image-finite and moreover the image of the transitive closure of $\mathbin{\totau}$
    is finite.
  \end{defi}

  The Hennessy-Milner theorem for branching bisimulation and modal logic
  was found by De Nicola and Vaandrager in~\cite{NicolaVaandrager}.
  The modal logic in question is Hennessy-Milner Logic with Until,
  where a diamond formula can impose a requirement not only on the state
  after the step, but also on all states on the path leading up to this state.
  We now briefly recap this logic, using the semantics introduced by
  van Glabbeek~\cite{GlabbeekSpectrum}.

  \begin{defi}
    Formulas of Hennessy-Milner Logic with Until (HMLU) are given by the following inductive definition
    \[
      \varphi := \top \mid \neg \varphi
              \mid \varphi_1 \wedge \varphi_2 \mid \varphi_1 \tuple{\alpha} \varphi_2.
    \]
  \end{defi}

  We will use $\varphi_1 \vee \varphi_2$ as a shorthand for $\neg (\neg \varphi_1 \wedge \neg \varphi_2)$
  and $\tuple{\alpha} \varphi$ as a shorthand for $\top \tuple{\alpha} \varphi$.
  The modality binds tightly on the left and loosely on the right,
  so $\delta \wedge \gamma \tuple{\alpha} \varphi \wedge \psi$ reads as
  $\delta \wedge (\gamma \tuple{\alpha} (\varphi \wedge \psi))$,
  and $\delta \tuple{\alpha} \varphi \tuple{\beta} \psi$ as $\delta \tuple{\alpha} (\varphi \tuple{\beta} \psi)$.

  The semantics of a formula $\varphi$ in a state $s$ of an $\LTS$
  are as follows.
  We write $s \vDash \varphi$ to mean $\varphi$ holds in $s$,
  and we write $s \rtc^\varphi_\alpha s'$ to indicate that there exists a sequence
  $s = s_1 \to_\alpha \ldots \toalpha s_n = s'$ such that for all $i \le n$, $s_i \vDash \varphi$.

  \begin{defi}\label{def.interp}
    The relation $s \vDash \varphi$ is defined inductively by the following rules:
    $$\begin{array}{lclclcl}
      s \vDash \top &\Leftrightarrow &\text{always} &\qquad\qquad&      s \vDash \varphi_1 \wedge \varphi_2 &\Leftrightarrow &\text{$s \vDash \varphi_1$ and $s \vDash \varphi_2$}\\
      s \vDash \neg \varphi &\Leftrightarrow &s \not\vDash \varphi &&
      s \vDash \varphi_1 \tuple{\alpha} \varphi_2 &\Leftrightarrow
          &\exists s', s''.\, \satcommoneventual{s}{\varphi_1} \wedge s'' \vDash \varphi_2.
    \end{array}$$
    Given a state $s$ we define the \emph{theory of $s$}, $\Th(s)$ as $\compr{\varphi \in \mathrm{HMLU}}{s \vDash \varphi}$.
    We say that $\varphi$ is a \emph{distinguishing formula} for $s$ and $t$ in case $\varphi \in \Th(s) - \Th(t)$  or $\varphi \in\Th(s) - \Th(t)$.
  \end{defi}

  Our choice to require $s' \maybetoalpha s''$ in the case of a formula with a modality
  follows van Glabbeek et al.~\cite{GlabbeekLuttikTricka} (and has
  also already been considered in~\cite{GlabbeekSpectrum}). It
  deviates from the definition by Vaandrager and de
  Nicola~\cite{NicolaVaandrager}, where a special case $\alpha =\tau$
  is considered. It can be shown that both interpretations of the
  modality generate the same logical equivalence of states (where
  two states $s$ and $t$ are logical equivalent if they have the same
  theory, $\Th(s)=\Th(t)$).

  Let us turn our attention to the semantic side of the question.
 
  \begin{defi}\label{def.branbis}
    A symmetric relation $R$ is a \emph{branching bisimulation} if whenever $R(s, t)$ and $s \toalpha s'$,
    there exists an eventual $\alpha$-step $\commoneventual{t}$ such that
    $R(s,t')$ and $R(s',t'')$. 
    Two states $s$ and $t$ are branching bisimilar (notation $s \bBisim t$) if they are related
    by some branching bisimulation.
  \end{defi}

  The situation is illustrated on the left in Figure~\ref{fig.branbis}.

    {\sizeChoice
    \begin{figure}
    \begin{tabular}{llrr}
      \hspace*{.5cm}&
      \begin{tikzpicture}[>=stealth,node distance=2.5cm,auto]
        \node  (s)                         {$s$};
        \node  (t) [right of = s]          {$t$};
        \node  (s') [below of = s]          {$s'$};
        \node  (t') [below of = t]          {$t'$};
        \node  (t'') [below of = t']          {$t''$};
        
        \path[<->,>=To]
        (s)   edge node{$R$}    (t);
        \path[<->,>=To,dotted]
        (s)   edge node{$R$}    (t')
        (s')   edge node{$R$}    (t'');
        \path[->]
        (s)   edge    node[swap] {$\alpha$} (s');
        \path[->,dotted]     
        (t')  edge    node {$(\alpha)$}  (t'');
        \path[->>,dotted]
        (t)   edge    node {$\tau$}  (t');
    \end{tikzpicture}
&\hspace*{1.5cm}&
    \begin{tikzpicture}[>=stealth,node distance=2.5cm,auto]
        \node  (s)                         {$s$};
        \node  (t) [right of = s]          {$t$};
        \node  (s') [below of = s]          {$s'$};
        \node  (s'') [below of = s']          {$s''$};
        \node  (t') [below of = t]          {$t'$};
        \node  (t'') [below of = t']          {$t''$};
        
        \path[->,>=To]
        (s)   edge node{$R$}    (t);
        \path[->,dotted,>=To]
        (s')   edge node{$R$}    (t');
        \path[<->,dotted,>=To]
        (s'')   edge node{$R$}    (t'');
        \path[->>]     
        (s)  edge    node[swap] {$\tau$}  (s');
        \path[->]
        (s')   edge    node[swap] {$(\alpha)$} (s'');
        \path[->>,dotted]
        (t)   edge    node {$\tau$}  (t');
        \path[->,dotted]     
        (t')  edge    node {$(\alpha)$}  (t'');
    \end{tikzpicture}
    \end{tabular}
    
    \caption{Branching Bisimulation (left), Directed Branching Bisimulation (right)}\label{fig.branbis}
  \end{figure}
  }
\noindent 
  In other words, a branching bisimulation is a relation that
  transfers $\alpha$-steps into eventual $\alpha$-steps.  In this
  definition we again follow Basten~\cite{Basten} and van Glabbeek
  et al.~\cite{GlabbeekLuttikTricka}. Basten calls these relations a
  ``semi-branching bisimulation'' and shows that the notion of being
  ``semi-branching bisimilar'' is equivalent to the original notion
  of branching bisimilar that can be found in the work of De Nicola
  and Vaandrager~\cite{NicolaVaandrager} and van
  Glabbeek~\cite{GlabbeekSpectrum}. An important advantage of the
  notion of branching bisimulation of Definition~\ref{def.branbis}
  is that the composition of two branching bisimulations is again a
  branching bisimulation.

  De Nicola and Vaandrager~\cite{NicolaVaandrager} have shown that
  branching bisimulation and HMLU are related via a Hennessy-Milner
  theorem. This theorem also applies to the notion of branching
  bisimulation of Definition~\ref{def.branbis} and the interpretation
  of HMLU formulas of Definition~\ref{def.interp}.

  \begin{thmC}[\cite{NicolaVaandrager}] \label{thm.dNV}
    $s \bBisim t$ precisely when $\Th(s) = \Th(t)$. 
  \end{thmC}

  \subsection{Positive Hennessy-Milner Logic with Until}
  \label{ssec.PHMLU}

  As was the case with Hennessy-Milner Logic, Hennessy-Milner Logic with Until
  needs to be adapted to a positive variant to allow for non-trivial inclusions
  of theories.
  Unlike HML, however, simply requiring that all negations be under a modality
  is insufficient for our purposes.
  Instead, we require that all negations be on the right-hand side of a modality.
  The motivation for this is as follows.

  A positive formula expresses what eventual $\alpha$-steps can be taken from some state $s$.
  If $s$ is a state that contains a silent step to $t$, then every eventual $\alpha$-step
  from $t$ can also be taken from $s$.
  Hence any positive formula that holds in $t$ should also hold in $s$;
  that is, for any positive formula $\varphi$, we expect the following implication to hold:
  \begin{equation}
    \label{eqn:silent_transfer}
    \text{$s \totau t$ and $t \vDash \varphi$} \quad\Rightarrow\quad s \vDash \varphi.
  \end{equation}

  \begin{exa}
    There exist formulas where all negations are under a modality that nevertheless
    do not satisfy implication (\ref{eqn:silent_transfer}).
    Consider $\varphi = (\neg\tuple{a}\top)\tuple{b}\top$ in the following LTS.
    
     {\small
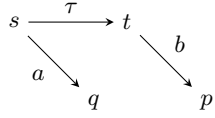
    \begin{tikzpicture}[>=stealth,node distance=1.5cm,auto]
      \node  (s)           {$s$};
      \node  (q) [below right of = s]                        {$q$};
      \node  (t) [right of = s]          {$t$};
      \node  (p) [below right of = t]          {$p$};
        \path[->]
          (t)   edge    node {$b$} (p)
          (s)  edge    node {$\tau$}  (t)
          (s)  edge    node[swap] {$a$}  (q);
    \end{tikzpicture}
    }

     We have $t \vDash \varphi$ but $s \not\vDash \varphi$.
  \end{exa}
  \noindent 
  To resolve this, we specify that a modality is only a positive formula if the formula
  on its left is positive, giving us the following definition.

  \begin{defi}
    We define \emph{positive} and \emph{negative} formulas by induction as follows:
    $$\begin{array}{lcllcl}
      \kwpos(\top) &\Leftrightarrow &\text{always}&
      \kwneg(\top) &\Leftrightarrow &\text{always}\\
      \kwpos(\neg \varphi) &\Leftrightarrow &\kwneg(\varphi)&
      \kwneg(\neg \varphi) &\Leftrightarrow &\kwpos(\varphi)\\
      \kwpos(\varphi_1 \wedge \varphi_2) &\Leftrightarrow &\text{$\kwpos(\varphi_1)$ and $\kwpos(\varphi_2)$}&
      \kwneg(\varphi_1 \wedge \varphi_2) &\Leftrightarrow &\text{$\kwneg(\varphi_1)$ and $\kwneg(\varphi_2)$}\\
      \kwpos(\varphi_1 \tuple{\alpha} \varphi_2) &\Leftrightarrow &\kwpos(\varphi_1)&
      \kwneg(\varphi_1 \tuple{\alpha} \varphi_2) &\Leftrightarrow &\text{never.}
    \end{array}$$
  \end{defi}
\noindent 
  Note that all modality-free formulas are both positive and negative.

  Positive and negative formulas satisfy the kind of transfer property we express in
  (\ref{eqn:silent_transfer}).

  \begin{lem}
    \label{lem.transfer}
    For $s, t$ states with $s \totau t$ we have
    \begin{itemize}
      \item for every positive formula $\varphi$, if $t \vDash \varphi$ then $s \vDash \varphi$;
      \item for every negative formula $\varphi$, if $s \vDash \varphi$ then $t \vDash \varphi$.
    \end{itemize}
  \end{lem}

  \begin{proof}
    By induction on the formula.
    The case when $\varphi = \delta \tuple{\alpha} \psi$ goes as follows: if $t \vDash \delta \tuple{\alpha} \psi$
    then there exists some $\satcommoneventual{t}{\delta}$.
    Since $\delta$ is positive and $t \vDash \delta$ we have $s \vDash \delta$ by induction.
    The path $s \totau \satcommoneventual{t}{\delta}$ witnesses that $s \vDash \delta \tuple{\alpha} \psi$,
    as desired.
  \end{proof}

  A class of formulas of some interest is the following:

  \begin{defi}
    A formula $\varphi$ is \emph{left-positive} if in all of its subformulas of the form $\delta \tuple{\alpha} \psi$,
    $\delta$ is positive.
  \end{defi}

  Note that not every left-positive formula is positive,
  and not every positive formula is left-positive.
  For example, $\neg (\top \tuple{\alpha} \top)$ is left-positive,
  since the formula $\top$ on the left of the modality is positive,
  but is as a whole not positive.
  On the other hand, $\tuple{\alpha} (\neg \tuple{\gamma} \top) \tuple{\beta} \top$
  is positive, but is not left-positive since $\neg \tuple{\beta} \top$ appears on
  the left of a modality but is not positive.\footnote{Recall our convention that $\tuple{\alpha} \varphi$
  is a shorthand for $\top \tuple{\alpha} \varphi$, and that modality binds strongly on the left and
  weakly on the right, so $\delta \tuple{\alpha} \varphi \tuple{\beta} \psi$ should be read as
  $\delta \tuple{\alpha} (\varphi \tuple{\beta} \psi)$.}
  As we shall see in Corollary~\ref{cor:lp_equiv},
  every HMLU formula is equivalent to a left-positive formula.
  Our experience is that proofs about left-positive formulas are simpler than
  about arbitrary formulas.
  This is largely due to the following simplification of the semantics:
  
  \begin{lem}
    \label{lem.positive_semantics}
    When restricted to left-positive formulas,
    the satisfaction relation is equivalent to the one generated by the following inductive definition:
    $$\begin{array}{lclclcl}
      s \vDash \top &\Leftrightarrow &\text{always}& &       s \vDash \varphi_1 \wedge \varphi_2 &\Leftrightarrow &\text{$s \vDash \varphi_1$ and $s \vDash \varphi_2$}\\
      s \vDash \neg \varphi &\Leftrightarrow &s \not\vDash \varphi &&
      s \vDash \varphi_1 \tuple{\alpha} \varphi_2 &\Leftrightarrow
          &\exists s', s''.\, \commoneventual{s} \wedge s' \vDash \varphi_1 \wedge s'' \vDash \varphi_2.
    \end{array}$$
  \end{lem}

  \begin{proof}
    By induction on the formula, using Lemma~\ref{lem.transfer}.
  \end{proof}

  In other words, for left-positive formulas the satisfaction relations of the \emph{until}
  modality and the \emph{just-before} modality coincide.
  The just-before modality has been studied amongst others by van Glabbeek~\cite{GlabbeekSpectrum},
  who remarks that both until and just-before give logics equivalent to branching bisimulation.
  The standard proof, given for example in~\cite{GlabbeekLuttikTricka}\footnote{This presentation
  features an explicit divergence operator, but this does not change the key argument.},
  is similar to the one we take here: one defines a class of upwards and
  downwards formulas and show that formulas of logic with until can
  be expressed as combinations of those.

  We take this approach a step further by defining a logic of positive formulas,
  giving them a syntactic characterisation.
  We call this logic Positive Hennessy-Milner Logic with Until, and ensure
  by construction that all formulas are both positive and left-positive.

  \begin{defi}
    Formulas of Positive Hennessy-Milner Logic with Until (PHMLU) are given by the following inductive definition
    \[
      \varphi := \top \mid \bot
              \mid \varphi_1 \wedge \varphi_2 \mid \varphi_1 \vee \varphi_2
              \mid \varphi_1 \tuple{\alpha} (\varphi_2 \wedge \neg \varphi_3).
    \]
  \end{defi}

  We again abbreviate $\top\tuple{\alpha}\varphi$ to $\tuple{\alpha}\varphi$
  and define $\ThP(s) := \compr{\varphi \in \PHMLU}{s \vDash \varphi}$.
  We regard formulas of PHMLU as also being formulas of HMLU, allowing us to use
  the same semantics.
  By induction, we see that all PHMLU formulas are positive and left-positive;
  in particular, this means that the transfer property of Lemma~\ref{lem.transfer}
  and the semantics of Lemma~\ref{lem.positive_semantics} hold for PHMLU formulas.

  Like with PHML (see Notation \ref{nota.phml}), there are two syntactic
  classes related to PHMLU that will be useful for our proofs,
  as well as some convenient shorthands.

  \begin{nota}\label{nota.phmlu}\hfill
    \begin{itemize}
      \item $\pmPHMLU$, the set of PHMLU formulas and their negations.
      \item $\cpmPHMLU$, the set of conjunctions of $\pmPHMLU$ formulas.
      \item Given a formula $\varphi \in \cpmPHMLU$, we again use $\varphi^+$ and $\varphi^-$
        to denote the PHMLU formulas such that $\varphi$ is equivalent to $\varphi^+ \wedge \neg \varphi^-$.
      \item Given formulas $\delta \in \PHMLU$ and $\varphi \in \cpmPHMLU$, we use $\delta \tuple{\alpha} \varphi$
        to denote $\delta \tuple{\alpha} \varphi^+ \wedge \neg \varphi^-$.
    \end{itemize}
  \end{nota}
\noindent 
  At this point, it may be concerning that we have both `positive HMLU formulas' and
  `PHMLU formulas', where the former is a superset of the latter; for example
  $\tuple{a}(\neg (\tuple{b} \top) \vee \tuple{a} \top)$ is positive but is not a PHMLU formula.
  The following theorem shows that when reasoning up to equivalence, this is not a problem,
  as every positive HMLU formula is equivalent to some PHMLU formula.

  \begin{thm}
    \label{thm.branching_theory_equiv}
    Every HMLU formula $\varphi$ is equivalent to a disjunction of $\cpmPHMLU$ formulas.
    Moreover, if $\varphi$ is positive then it is equivalent to a disjunction of $\PHMLU$ formulas.
  \end{thm}

  The full proof can be found in Appendix~\ref{app.proofs}.
  Here, we restrict ourselves to a proof sketch.

  \begin{proof}[Proof sketch]
    The proof is similar to that of Theorem~\ref{thm.strong_theory_equiv}.  
    We again show, by induction, that every formula $\varphi$
    is equivalent to a disjunction $\bigvee_{i \in I} \Phi_i$,
    and only the modality case is of interest.

    Let $\delta \tuple{\alpha} \psi$ be an HMLU formula.
    By induction, $\delta$ is equivalent to $\check{\Delta} = \bigvee_i \Delta_i$ and
    $\psi$ is equivalent to $\bigvee_j \Psi_j$.
    Since the modality distributes over disjunction on the right,
    we can take the same approach as in Theorem~\ref{thm.strong_theory_equiv}
    to show that $\delta\tuple{\alpha}\psi$ is equivalent to
    $\bigvee_j \left(\check{\Delta} \tuple{\alpha} \Psi_j\right)$.
    It remains to show that for every $j$, the disjunct $\check{\Delta} \tuple{\alpha} \Psi_j$
    is equivalent to a disjunction of $\cpmPHMLU$ formulas.

    For any state $s$, if $s \vDash \check{\Delta} \tuple{\alpha} \Psi_j$
    then there is some path $\satcommoneventual{s}{\check{\Delta}}$ with $s''\vDash \Psi_j$.
    By taking the path $s = s_1 \totau s_2 \totau \ldots \totau s_n$ and finding the disjunct
    $\Delta_{i_k}$ that holds in state $s_k$, we can find a formula
    \begin{equation}\label{eqn:bad_phmlu_equiv}
      \Delta_{i_1} \tuple{\tau} \Delta_{i_2} \tuple{\tau} \ldots \tuple{\tau} \Delta_{i_n} \tuple{\alpha} \Psi_j.
    \end{equation}
    
    This formula holds in $s$, and conversely, if any formula of this shape holds in $s$,
    then so does $\check{\Delta} \tuple{\alpha} \Psi_j$.
    However, we cannot take the disjunction of all such formulas as our solution:
    these formulas are not necessarily $\cpmPHMLU$,
    and there may be infinitely many of them.

    To remedy this, note that for any $\cpmPHMLU$ formula $\theta$,
    the formula $\Delta_{i_k} \tuple{\tau} \theta$
    is equivalent to $\neg \Delta_{i_k}^- \wedge \Delta_{i_k}^+ \tuple{\tau} \theta$.
    This is because by Lemma~\ref{lem.transfer}, satisfaction of
    negative formulas is closed under silent steps,
    so if $\neg\Delta^-_{i_k}$ holds in some state $r$, it will hold in all states reachable from $r$ by silent steps.
    The formula (\ref{eqn:bad_phmlu_equiv}) is thus equivalent to
    \[
      \neg\Delta_{i_1}^- \wedge \Delta_{i_1}^+ \tuple{\tau} \neg\Delta_{i_2}^- \wedge \Delta_{i_2}^+ \tuple{\tau} \ldots \tuple{\tau} \neg\Delta_{i_n}^- \wedge \Delta_{i_n}^+ \tuple{\alpha} \Psi_j.
    \]

    Recall that by our convention, the modality binds tightly on the left and loosely on the right,
    so this should be read as $\neg\Delta_{i_1}^- \wedge (\Delta_{i_1}^+ \tuple{\tau} (\ldots))$.

    Moreover, suppose that for some $l, k$, $i_l = i_k$.
    Then in fact $s_{k'} \vDash \Delta_{i_k}$ for all $k'$ between $k$ and $l$.
    Namely, by Lemma~\ref{lem.transfer},
    all the positive formulas of $\Delta_{i_l}$ will hold for $s_{k'}$ if $k' \le l$,
    while all the negative formulas of $\Delta_{i_k}$ will hold for $s_{k'}$ if $k' \ge k$.
    It follows that a single modality can be used for all the steps from $l$ to $k$,
    meaning that the number of extra $\tau$ modalities we use in the construction
    of our formula can be bounded by the number of disjuncts in $\Delta$ (in fact, minus one).

    Putting this together, we get the following disjunction,
    where $i_1, \ldots, i_n$ ranges over all non-empty sequences of indices in $I$ without doubles. 
    \[
      \Phi := \bigvee_{\vec i, j} \neg\Delta_{i_1}^- \wedge \Delta_{i_1}^+ \tuple{\tau} \ldots \tuple{\tau} \neg\Delta_{i_n}^- \wedge \Delta_{i_n}^+ \tuple{\alpha} \Psi_j.
    \]

    This concludes our sketch of the construction of the equivalent formula.
    A more explicit construction, including its verification, can be found in Appendix~\ref{app.proofs}.

    The second part of the theorem, showing that this construction can be
    done without negations of PHMLU formulas when $\varphi$ is positive,
    again goes by induction: in this case $\Delta_{i_1}^- = \bot$,
    from which it is immediate that each disjunct of $\Phi$ is,
    up to equivalence, a PHMLU formula.
  \end{proof}

  \begin{cor}
    Every positive formula has an equivalent PHMLU formula.
  \end{cor}

  \begin{proof}
    Immediate: a disjunction of PHMLU formulas is itself a PHMLU formula.
  \end{proof}

  \begin{cor}
    \label{cor:lp_equiv}
    Every formula has an equivalent left-positive formula.
  \end{cor}
  
  \begin{proof}
    Immediate: every Boolean combination of PHMLU formulas is left-positive.
  \end{proof}

  \begin{cor}
    For two states $s, t$, $\ThP(s) = \ThP(t)$ if and only if $\Th(s) = \Th(t)$.
  \end{cor}

  \begin{proof}
    This proof is essentially the same as the proof of Corollary~\ref{cor:strong_theory_equality}.
  \end{proof}

  \subsection{Directed branching bisimulation}
  \label{ssec.dirbranbis}

  While branching bisimulation takes $\alpha$-steps to eventual $\alpha$-steps,
  we let directed branching bisimulation take eventual $\alpha$-steps to
  eventual $\alpha$-steps.
  We drop the symmetry requirement on the relation itself, instead requiring the
  final states in these steps to be related in both directions.
  This is what makes our notion a directed bisimulation, rather than simply a
  simulation.

  \begin{defi}
    A relation $R$ is a \emph{directed branching bisimulation} if whenever $R(s,t)$ and $\commoneventual{s}$,
    there exist $t', t''$ such that $\commoneventual{t}$, $R(s',t')$, $R(s'',t'')$, and $R(t'',s'')$.
    See the right side of Figure~\ref{fig.branbis}.

    Two states are \emph{directed branching bisimilar} (notation $s \dBisim t$) if there exists a directed branching bisimulation
    relating them.
    We write $s \dBisimSym t$ as a shorthand for ``$s \dBisim t$ and $t \dBisim s$''.
  \end{defi}

  A number of properties that are known for branching bisimulation have simple proofs in the directed setting.
  In particular, the stuttering lemma can be stated in the following, more specific, way.

  \begin{lem}[Stuttering lemma] See Figure~\ref{fig.stut} on the left.
    \label{lem.bisim_stuttering}
    Given states $s \dBisim t$, we have for all $q, r$,
    \begin{itemize}
      \item if $s \rtctau q$ then $q \dBisim t$, and
      \item if $r \rtctau t$ then $s \dBisim r$.
    \end{itemize}
  \end{lem}
\noindent 
  This can be proven using a stuttering closure construction like the one used by van Glabbeek et al.~\cite{GlabbeekLuttikTricka},
  but we will instead prove its dual, Lemma~\ref{lem.apart_stuttering}.

  \begin{thm}
    \label{thm.directed_branching_symm_interior}
    Two states are branching bisimilar precisely when they are directed branching bisimilar in both directions:
    $$ s\bBisim t \qquad\Longleftrightarrow \qquad s \dBisim t \wedge t \dBisim s$$
  \end{thm}

  \begin{proof}
    Every branching bisimilarity is a directed branching bisimilarity, giving the left to right direction.

    For the other direction, it suffices to show that the relation $\mathbin{\dBisimSym}$
    is a branching bisimilarity.
    This is easily verified: it is symmetric, and given $s, s', t$ such that $s \toalpha s'$
    and $s \dBisimSym t$, there exist some $t', t''$ such that $\commoneventual{t}$
    with $s \dBisim t'$ and $s' \dBisimSym t''$.
    It remains to show that $t' \dBisim s$: this follows from the Stuttering Lemma, since
    $t \dBisim s$ and $t \rtctau t'$.
  \end{proof}

  \subsection{Directed branching apartness}
  \label{ssec.dirbranapt}
  In the previous two subsections, we have introduced Positive Hennessy-Milner Logic with Until
  and directed branching bisimulation, which together are enough to formulate a directed
  Hennessy-Milner theorem.
  In this subsection, we introduce directed branching apartness, the dual relation of directed branching bisimulation.
  Using apartness makes the proof of our directed Hennessy-Milner theorem a straightforward
  induction in both directions.

  \begin{defi}
    A relation $Q$ is a \emph{directed branching apartness} if the following derivation rule holds:
    \[
      \inferrule*[Right=$\textsc{in}_{db}$]{
        \commoneventual{s} \\
        \forall t', t''.\, \commoneventual{t} \implies Q(s', t') \vee Q(s'', t'') \vee Q(t'', s'')
      }{Q(s, t)}
    \]

    Two states $s, t$ are \emph{directed branching apart}, notation $s
    \dApart t$, if for every directed branching apartness $Q$, it
    holds that $Q(s, t)$.  We use $s \dApartSym t$ as a shorthand for
    ``$s \dApart t$ or $t \dApart s$''.
  \end{defi}

  Just as for directed strong apartness (Definition~\ref{def.dirapt}),
  directed branching apartness is an {\em inductive\/} notion,
  so it can be characterized as a derivation system.
  For directed branching apartness,
  we have a lemma similar to Lemma~\ref{lem.dirsapt}.
  The proof is straightforward.
  
  \begin{lem} \label{lem.dirbapt}
    We have that $s \dApart t$ if and only this can be derived using the following rule.
      \[
        \inferrule*[Right=$\textsc{in}_{db}$]{
          \commoneventual{s} \\
          \forall t', t''.\, \commoneventual{t} \implies s' \dApart t' \vee s'' \dApartSym t''
        }{s \dApart t}
      \]
    Furthermore, $s \dApart t \quad\Leftrightarrow \quad \neg (s\dBisim t)$.
  \end{lem}

  We again note, as we did after Lemma~\ref{lem.dirsapt}, that
  the base case of the inductive definition of $s\dApart t$
  is when $s$ can do an eventual $\alpha$-step while $t$ cannot.
  Then the hypothesis is vacuously satisfied and we have $s\dApart t$.
    
  For all $s, t$ we have either $s \dBisim t$ or $s \dApart
  t$, depending on whether there is any directed branching bisimulation
  that relates them. So we can obtain results about $\dApart$ from $\dBisim$ (and vice versa) by contraposition.
  The following lemma is the Stuttering Lemma for apartness,
  the dual of Lemma~\ref{lem.bisim_stuttering}.

  \begin{lem}
    \label{lem.apart_stuttering}
    Given states $s \dApart t$, we have for all $r, q$,
    \begin{itemize}
      \item if $r \rtctau s$ then $r \dApart t$; and
      \item if $t \rtctau q$ then $s \dApart q$.
    \end{itemize}
    See the right side of Figure~\ref{fig.stut}.
  \end{lem}

  \begin{figure}
\begin{tabular}{lcr}
      \begin{tikzpicture}[>=stealth,node distance=2.5cm,auto]
        \node  (s)                         {$s$};
        \node  (t) [right of = s]          {$t$};
        \node  (q) [below of = s]          {$q$};
        \node  (r) [above of = t]          {$r$};
        
        \path[->,>=To]
        (s)   edge node{$_{db}$}    (t);
        \path[->,>=To,dotted]
        (s)   edge node{$_{db}$}    (r)
        (q)   edge node[swap] {$^{db}$}    (t);
        \path[->>]
        (s)   edge    node {$\tau$}  (q)
        (r)   edge    node {$\tau$}  (t);    
    \end{tikzpicture}
    &\hspace*{2cm}&
    \begin{tikzpicture}[>=stealth,node distance=2.5cm,auto]
        \node  (s)                         {$s$};
        \node  (t) [right of = s]          {$t$};
        \node  (q) [below of = t]          {$q$};
        \node  (r) [above of = s]          {$r$};
        
        \path[->]
        (s)   edge node{$_{\neg db}$}    (t);
        \path[->,>=To,dotted]
        (s)   edge node[swap]{$^{\neg db}$}    (q)
        (r)   edge node {$_{\neg db}$}    (t);
        \path[->>]
        (t)   edge    node {$\tau$}  (q)
        (r)   edge    node[swap] {$\tau$}  (s);    
    \end{tikzpicture}
\end{tabular}
    \caption{Stuttering using directed branching bisimilarity (left) and directed branching apartness (right)}\label{fig.stut}
  \end{figure}

  \begin{proof}
    Since $s \dApart t$, there exists (by Lemma~\ref{lem.dirbapt})
    some eventual $\alpha$-step $\commoneventual{s}$ such that
    for all eventual $\alpha$-steps $\commoneventual{t}$ we have
    $s' \dApart t'$ or $s'' \dApartSym t''$.

    Given $r \rtctau s$, we can use the same eventual $\alpha$-step $r \rtctau \commoneventual{s}$
    to show that $r \dApart t$, since the condition on eventual $\alpha$-steps from $t$ does not mention $s$ itself.

\vspace{0.5\baselineskip} 
    Similarly, given $t \rtctau q$, any eventual $\alpha$-step from $q$ is in fact an eventual $\alpha$-step from $t$.
    Hence for any such step $\commoneventual{q}$ we have $s' \dApart q'$ or $s'' \dApartSym q''$, as desired.
  \end{proof}

  We now state a theorem similar to Theorem~\ref{thm.directed_strong_apt_hm} for the
  branching case: $s \dsApart t$ iff there is some PHMLU formula that
  distinguishes $s$ and $t$ (from which the well-known Hennessy-Milner
  theorem for branching bisimulation of~\cite{NicolaVaandrager}, Theorem \ref{thm.dNV},
  immediately follows). Moreover we have a direct proof by induction.

  \begin{thm}
    \label{thm.directed_branching_hm}
    In an image-finite labelled transition system we have, for states $s$ and
    $t$,
    $$s \dApart t \quad\Longleftrightarrow \quad \exists \varphi.\, \varphi \in \ThP(s) - \ThP(t).$$
    Equivalently, $s$ and $t$ are directed branching bisimilar, $s \dBisim t$,
    precisely when $\ThP(s) \subseteq \ThP(t)$.
  \end{thm}

  \begin{proof}
    We only prove the first statement. Both directions of the proof proceed by structural induction.
    
    $\Rightarrow$: Given a derivation of $s \dApart t$, we know it was obtained by applying derivation rule $\textsc{IN}_{db}$,
    and hence there exist $s', s''$ such that $\commoneventual{s}$ such that for all
    $t', t''$, if $\commoneventual{t}$ then $s' \dApart t'$, $s'' \dApart t''$ or $t'' \dApart s''$.
    By induction, for each of these cases there is a distinguishing formula, which we can
    collect into sets $\Delta$, $\Psi^+$, and $\Psi^-$.
    Let $\delta = \bigwedge \Delta$, $\psi^+ = \bigwedge \Psi^+$, and $\psi^- = \bigvee \Psi^-$.
    We have that $s' \vDash \delta$, $s'' \vDash \psi^+$, and $s'' \not\vDash \psi^-$,
    while for all $t', t''$ with $\commoneventual{t}$, we have
    $t' \not\vDash \delta$, $t'' \not\vDash \psi^+$, or $t'' \vDash \psi^-$.
    Let $\phi = \delta\tuple{\alpha}\psi^+ \wedge \neg \psi^-$;
    by construction, $s \vDash \phi$ while $t \not\vDash \phi$, hence $\phi$ is a positive distinguishing formula.

    $\Leftarrow$: In this direction, only the modality case is of interest.
    Suppose $\varphi = \delta \tuple{\alpha} \psi^+ \wedge \neg \psi^-$ is a
    positive distinguishing formula for $s$ and $t$.
    Then there exist $s', s''$ such that $\commoneventual{s}$,
    such that $s' \vDash \delta$, $s'' \vDash \psi^+$, and $s'' \not\vDash \psi^-$,
    while for all $t', t''$ such that $\commoneventual{t}$
    one of $t' \not\vDash \delta$, $t'' \not\vDash \psi^+$, and $t'' \vDash \psi^-$ holds.
    By induction, we can conclude $s' \dApart t'$, $s'' \dApart t''$, or $t'' \dApart s''$,
    which is exactly what we need to derive the apartness $s \dApart t$.
  \end{proof}

The proof of Theorem~\ref{thm.directed_branching_hm} gives an
algorithm to generate a distinguishing PHMLU formula from a
proof of directed apartness and vice versa.
It extends the algorithms for the  strong and weak bisimulation case as described in \cite{Geuvers}.
We illustrate this using a well-known LTS with states
$s$ and $r$ that are not branching bisimilar.
We give the proof of $s\dApart r$ and the corresponding PHMLU formula
that distinguishes $s$ and $t$.
In Appendix~\ref{app.example} we describe the algorithm and
give an example of the reverse direction.
  
\begin{exa}\label{exa.LTSbranapt}
In the following well-known LTS, $s$ and $r$ are not
branching bisimilar.  We give a proof of $s\dApart r$ and the
distinguishing formula that is computed from that proof.
{\sizeChoice
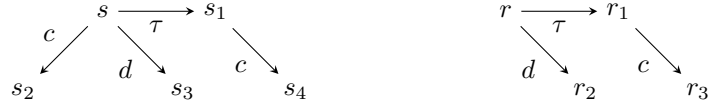
  \begin{center}
  \begin{tabular}[t]{rcl}
    \begin{tikzpicture}[>=stealth,node distance=1.5cm,auto]
        \node    (s)                       {$s$};
        \node    (s1) [right of = s]  {$s_1$};
        \node    (s4) [below right of = s1]      {$s_4$};
        \node    (s3) [below right of = s] {$s_3$};
        \node    (s2) [below left of = s]       {$s_2$};

        \path[->]
            (s)  edge                    node[swap] {$\tau$}  (s1)
                edge                    node[swap] {$d$}     (s3)
                edge                    node[swap] {$c$}     (s2)
            (s1) edge                    node[swap] {$c$}     (s4);
    \end{tikzpicture}
    &
    \qquad\qquad
    &
    \begin{tikzpicture}[>=stealth,node distance=1.5cm,auto]
        \node   (v)  [right of = s1]                     {};
        \node   (r)  [right of = v]                     {$r$};                
        \node   (r2) [below right of = r] {$r_2$};
        \node   (r1) [right of = r]  {$r_1$};
        \node   (r3) [below right of = r1]      {$r_3$};

        \path[->]
            (r)  edge                    node[swap] {$\tau$}  (r1)
                 edge                    node[swap] {$d$}     (r2)
            (r1) edge                    node[swap] {$c$}     (r3);
    \end{tikzpicture}
  \end{tabular}
\end{center}
}
In the derivation of $s\dApart r$, we indicate above the
$\forall r', r'' \ldots$ all the (finitely many) possible transitions
that we need to prove a property for: just one in the case of the
$c$-step; none in the case of the $d$-step (which we indicate by $\checkmark$).
{\sizeChoice
  \[
  \inferrule*{ s \to_c s_2 \quad
     \inferrule*{
        \inferrule*{
           \inferrule*{ s \to_d s_3 \quad
             \inferrule*{\checkmark
                        }
                        {\forall r', r''.\, r_1 \rtctau r'\to_d r'' \implies s \dApart r' \vee s_3 \dApartSym r''
                        }
                      }
                      {s \dApart r_1
                      }
                   }
               {s \dApart r_1 \vee s_2 \dApartSym r_3}
                }
             {\forall r', r''.\, r \rtctau r'\to_c r'' \implies s \dApart r' \vee s_2 \dApartSym r''
               }
  }
    {s\dApart r}
\]
}
The formula that we compute from this proof is
$(\tuple{d}\top)\tuple{c} \top$, and indeed  $s \vDash (\tuple{d}\top)\tuple{c} \top$ and $r \not\vDash (\tuple{d}\top)\tuple{c} \top$.
The formula  expresses that there is a $\tau$-path to a state where a $c$-step is
possible, and in all states along that $\tau$-path, a $d$-step is
possible.
\end{exa}

Using the Stuttering Lemma~\ref{lem.bisim_stuttering} we can prove the
following interesting alternative characterization of directed
branching bisimulation.
  
  \begin{thm}
    \label{thm.branching_bisim_char}
    For states $s, t$, we have $s \dBisim t$ precisely when there exists some $r$ such that $t \rtctau r$ and $s \dBisimSym r$.
  \end{thm}

  \begin{proof}
    Suppose $s \dBisim t$. Since $\eventual{\tau}{s}{s}{s}$ is an eventual $\tau$-step,
    there exist $t'$ and $t''$ such that $\eventual{\tau}{t}{t'}{t''}$,
    $s \dBisim t'$, and $s \dBisimSym t''$.
    For the reverse implication, suppose $t \rtctau r$ and $s \dBisimSym r$.
    Then also  $s \dBisim r$, so by the Stuttering Lemma~\ref{lem.bisim_stuttering}, we have $s \dBisim t$.
  \end{proof}

  This characterization, together with Theorem~\ref{thm.directed_branching_hm}, gives us the following corollary.

  \begin{cor}\label{cor:incl_char}
    Given two states $s, t$ in an image-finite $\LTStau$, $\ThP(s) \subseteq \ThP(t)$
    precisely when there exists some $t'$ such that $t \rtctau t'$ and $\ThP(s) = \ThP(t')$. 
  \end{cor}
  \noindent 
  This result is particularly interesting because it does not mention bisimulation or apartness,
  but it is unclear how one would go about proving it purely within (P)HMLU.
  While for every formula $\varphi \in \ThP(s) - \ThP(t)$ we know $t \vDash \tuple{\tau} \neg \varphi$ and hence there is
  some $s'$ such that $s \rtctau s'$ and $s' \not\vDash \varphi$, we see no way to conclude that a single $s'$
  will work independent of $\varphi$.
  The equivalence with directed branching bisimulation, however, gives us exactly this.

  \subsection{Relation to other notions}

  In the non-directed case, strong bisimulation implies branching bisimulation,
  so it may come as a surprise that this does not hold in the directed case.
  This can be seen as follows: a state with no outgoing transitions is directed strongly
  bisimilar to every other state, since the condition that every transition can
  be transferred is satisfied vacuously.
  In other words, its positive theory contains no formulas containing modalities,
  and it is thus minimal with respect to inclusion.
  However, in the branching case a state with no outgoing transitions satisfies
  formulas such as $\tuple{\tau}\neg \tuple{a}\top$, which are not true in every
  state.
  
  A similar surprise can be found by comparing directed branching bisimulation
  to the following notion introduced by Bergstra and Klop in~\cite{BergstraKlop},
  who are working in the context of process algebras and have an addition operation
  on states.
  In our terms, the state $s + t$ should be seen as the state that can do anything $s$ or $t$
  can do; in other words, $s + t \to_a r$ precisely when $s \to_a r$ or $t \to_a r$.
  Bergstra and Klop define $s \sqsubseteq t$ as follows: $s = t$ or there exists a $r$ such that $s + r = t$.
  In our context, it makes most sense to interpret this equality as bisimulation.

  If we treat the equality as strong bisimulation, we in fact have that $s
  \dsBisim t$ precisely when $s \sqsubseteq t$.
  However, in the branching case this does not hold: if $s$ is the state with no
  outgoing transitions, then $s \sqsubseteq t$ for every $t$, but $s \dBisim t$
  does not in general hold.
  A reviewer also pointed out the following example, which shows that
  there exist states $u$ and $v$ such that $u \dBisim v$ but not $u + v \dBisim v$,
  and hence that directed branching bisimulation is not well-behaved under such
  operation on states:

{\sizeChoice
  \begin{center}
  \begin{tabular}[t]{rclcl}
    \begin{tikzpicture}[>=stealth,node distance=1.5cm,auto]
        \node   (s)  []                     {$u$};
        \node   (s1) [below left of = s]   {$\cdot$};

        \path[->]
            (s) edge                    node[swap] {$a$}     (s1);
    \end{tikzpicture}
    &
    \qquad\qquad
    &
    \begin{tikzpicture}[>=stealth,node distance=1.5cm,auto]
        \node   (t)  []                     {$v$};
        \node   (t1) [right of = t]          {$\cdot$};
        \node   (t2) [below left of = t1]    {$\cdot$};
        \node   (t3) [below right of = t1]   {$\cdot$};

        \path[->]
            (t)  edge                    node {$\tau$}(t1)
            (t1) edge                    node {$a$}    (t2)
                 edge                    node[swap] {$b$}    (t3);
    \end{tikzpicture}
    &
    \qquad\qquad
    &
    \begin{tikzpicture}[>=stealth,node distance=1.5cm,auto]
        \node    (st)                       {$u+v$};
        \node    (st1) [right of = st]  {$\cdot$};
        \node    (st4) [below right of = st1]      {$\cdot$};
        \node    (st3) [below left of = st1] {$\cdot$};
        \node    (st2) [below left of = st]       {$\cdot$};

        \path[->]
            (st)  edge                    node {$\tau$}  (st1)
                edge                    node[swap] {$a$}     (st2)
            (st1)    edge                    node {$a$}     (st3)
                  edge                    node[swap] {$b$}     (st4);
    \end{tikzpicture}
  \end{tabular}
\end{center}
}

  Both the comparison to directed strong bisimulation and to the $\sqsubseteq$ relation
  raise the question: why does directed branching bisimulation not relate states $s$
  and $t$ even though $t$ appears to be able to do more than $s$ can?

  As we will show in a moment, the relation $\dsbBisim = {\dsBisim} \cup
  {\dBisim}$ also satisfies the property that $p \bBisim q$ precisely when $p
  \dsbBisim q \wedge q \dsbBisim p$.\footnote{
    Strictly speaking, we did not define directed strong bisimulation for an LTS with silent steps.
    We interpret $\tau$ as being no different than any other step for the purposes of directed strong
    bisimulation.
  }
  In this sense, we could weaken the relation $\dBisim$ to obtain one that behaves well with
  respect to $\sqsubseteq$.
  However, our result is not simply the existence of the $\dBisim$ relation:
  we show that it can be defined coinductively, that it has an inductive complement,
  and prove a Hennessy-Milner theorem that shows that $p$ and $q$ being directed branching
  bisimilar is equivalent to the PHMLU theory of $p$ being a subset of that of $q$,
  which in turn is equivalent to every positive and left-positive HMLU formula
  that holds in $p$ also holding in $q$.

  Finding a coinductive presentation and a corresponding logic for the relation
  $\dsbBisim$ (or for some other relation ${\dsbBisim} \subseteq R \subset {\bBisim}$)
  would be interesting future work.
  However, such work would not invalidate or supersede ours: the Hennesy-Milner theorem for
  a (coinductively defined) $\dsbBisim$ and some modal logic would simply show a connection
  between those two; the connection between $\dBisim$ and PHMLU is independent of that.

  \begin{thm}
    In any $\LTStau$, for all $p, q$ we have $p \bBisim q$ iff $p \dsbBisim q$ and $q \dsbBisim p$.
  \end{thm}

  \begin{proof}
    If $p \bBisim q$ then by Theorem~\ref{thm.directed_branching_symm_interior} we have $p \dBisim q$
    and $q \dBisim p$, which suffices.

    For the other direction,
    the only interesting case is when (without loss of generality)
    $p \dsBisim q$ and $q \dBisim p$, since strong bisimilarity implies branching bisimilarity
    in the undirected case.

    We show that in this case, $p \dBisim q$ by showing that $\ThP(p) \subseteq \ThP(q)$
    by induction on the formula.
    As in other proofs, only the case when the formula is of the form $\delta \tuple{\alpha} \psi$
    is interesting.

    Suppose that $p \vDash \delta \tuple{\alpha} \psi$, then there is some path
    $\satcommoneventual{p}{\delta}$ such that $p'' \vDash \psi$.
    By induction, $q \vDash \delta$.
    If this path has at least one step then we are done: if $p \neq p'$
    then by repeated applications of directed strong bisimilarity
    we can construct a path $q \rtctau^\delta q'$ such that $p' \sBisim q'$
    and hence $p' \bBisim q'$.
    By our choice of $p'$ we have $p' \vDash \delta \tuple{\alpha} \psi$
    and hence by bisimilarity $q' \vDash \delta \tuple{\alpha} \psi$,
    and by positivity $q \vDash \delta \tuple{\alpha} \psi$.
    Similarly, if $p = p'$ but $p' \neq p''$ then using directed strong
    bisimilarity we get $p'' \sBisim q''$ and thus $p'' \bBisim q''$.
    Since $p'' \vDash \psi$ also $q'' \vDash \psi$, and thus $q \vDash \delta \tuple{\alpha} \psi$.

    It remains to consider the case when the path is trivial, $p = p' = p''$, and thus $\alpha = \tau$.
    We use the fact we can write $\psi = \psi^+ \wedge \neg \psi^-$, where both 
    $\psi^+$ and $\psi^-$ are PHMLU formulas, and we have $p \vDash \psi^+$ and $p \not\vDash \psi^-$.
    Now since $\psi^+$ is a PHMLU formula, by our induction hypothesis we have $q \vDash \psi^+$.
    On the other hand, since $q \dBisim p$, every PHMLU formula that holds on $q$
    also holds in $p$, and by the contraposition of this we get $q \not\vDash \psi^-$.
    It follows that $q \vDash \delta \tuple{\alpha} \psi$ via a trivial path as well,
    which concludes the proof.
  \end{proof}
  
  \section{Related Work}

  Our approach for defining formulas is somewhat similar to that taken by 
  Beohar et al in~\cite{BeoharEtal}.
  There too, the `formulas' (elements of $\mathbb{L}$ in their terminology)
  are only closed under negation under the modality.
  This corresponds to our presentation of PHML in this paper.
  It would be interesting to see whether our results can be formulated in their setting.

  Our proof of Theorem~\ref{thm.branching_theory_equiv}, while discovered independently,
  is very similar to the proofs from Section~6 of van Glabbeek et al.~\cite{GlabbeekLuttikTricka},
  where the authors show that an `until' modality
  (denoted by `$\delta \tuple{\alpha} \psi$', where $s$ satisfies $\delta \tuple{\alpha} \psi$ when there is a path $\satcommoneventual{s}{\delta}$ with $s'' \vDash \psi$)
  is equally expressive as a `just-before' modality
  (denoted by `$\delta \,\alpha\, \psi$', where $s$ satisfies $\delta \,\alpha\, \psi$ when there is a path $\commoneventual{s}$ with $s' \vDash \delta$ and $s'' \vDash \psi$).
  Note that when $\delta$ is positive, this is trivial, since the same paths satisfy both conditions,
  and in fact this same approach is how van Glabbeek et al.\@ proved their results.

  The case of branching bisimulation and a modal logic with a `just-before' modality has
  been studied by Martens and Groote~\cite{MartensGroote}, who proved a Hennessy-Milner
  using an inductive argument on the apartness derivation, precisely of the sort that
  Geuvers~\cite{Geuvers} set out as an open question.
  Given the aforementioned equivalence in power between the `until' and `just-before' modalities,
  the two results imply each other.

  A notion similar to directed bisimulation is the one introduced by Bergstra and Klop in~\cite{BergstraKlop},
  who are working in the context of process algebras and have defined an addition operation.

  Moreover, even for strong bisimulation we feel that our presentation gives something that
  is not covered by the process algebra view.
  Namely, directed bisimulation is built up coinductively, and has an inductive dual.
  This means that the structure of a proof of $s \dsBisim t$ or $s \dsApart t$ can be useful
  when reasoning about a labelled transition system.
  For the same reason, we do not use the characterization in Theorem~\ref{thm.branching_bisim_char}
  as our definition of directed branching bisimulation.

  For Kripke structures, a notion of ``directed modal simulation'' has been
  studied by Kurtonina and de Rijke in \cite{KurtoninadeRijke}, where it is also
  related to theories of positive formulas in modal logic.
  Bisimulation for Kripke structures is comparable to strong bisimulation for
  LTSs, so silent steps are not treated.
  One of the goals of \cite{KurtoninadeRijke} is to capture the class of modal
  formulas that are preserved by a directed modal simulation, which includes
  box formulas, but does not include formulas where a negation appears under
  a modality.
  As a result, their directed modal simulation is incomparable to ours: it
  includes a condition for backwards transfer, which ours does not, but does not
  require that successor states be bisimilar.

  Notions of bisimilarity exist for a large variety of systems and
  calculi (process algebras, $\pi$-calculus, ...), for example see
  \cite{Sangiorgi}.  Desharnais et al~\cite{DesharnaisEtAl} have
  studied bisimulation relations on Markov processes, and in their
  work, bisimulation is equivalent to two-way simulation.  In
  particular, we would call their Proposition~2.11 a directed
  Hennessy-Milner theorem.
  
  \section{Conclusion}

  We have formulated and proved a directed Hennessy-Milner
  theorem, by adapting known Hennessy-Milner logic to a positive
  variant and by adapting bisimulation to directed bisimulation. From
  this directed version, the usual Hennessy-Milner theorem follows as
  a corollary. We also show that the dual of bisimulation,
  apartness (and its directed variant), 
  enables a direct proof of the Hennessy-Milner theorem by induction.

  We have illustrated this using LTSs (labelled transition systems) with strong
  bisimulation and using LTSs with silent steps with branching bisimulation.
  The situation for LTSs with silent steps and weak bisimulation is very similar to strong bisimulation.
  Applications of our approach to other logics and notions of bisimilarity would be of interest.
  Variations like divergence-preserving branching bisimulation~\cite{GlabbeekLuttikTricka,Luttik}
  would be interesting to consider from this directed perspective.

  In the branching setting, our ``directed approach'' brings the logic and bisimulation
  sides of the Hennessy Milner theorem closer together.
  There is a remarkable
  similarity between proving that a positive formula holds in a state
  and proving that a state is on the right of a directed bisimilarity.
  Namely, to show that $t \vDash \delta \tuple{\alpha} (\psi^+ \wedge
  \neg \psi^-)$ we show
  \[
    \exists t', t''.\, \commoneventual{t} \wedge t' \vDash \delta \wedge t'' \vDash \psi^+ \wedge t'' \not\vDash \psi^-,
  \]
  while to show that $s \dsBisim t$, we show that for every step $s \to_\alpha s'$,
  \[
    \exists t', t''.\, \commoneventual{t} \wedge s \dsBisim t' \wedge s' \dsBisim t'' \wedge t'' \dsBisim s'.
  \]
  Similarly, showing that a formula fails to hold in a state
  is akin to showing that the state is on the right of a directed apartness.

  This level of similarity suggests that there should be some way of constructing
  a modal logic from a coinductive notion of bisimilarity, or vice-versa,
  but this has so far proved elusive.
  For bisimulation relations, this question is raised in a considerably more general setting
  by Kupke and Rot~\cite{KupkeRot}.
  The work of Beohar et al~\cite{BeoharEtal} may be relevant to this point.

  \bibliographystyle{alphaurl}
  \bibliography{bibliography}

  \appendix

  \section{Proofs}
  \label{app.proofs}

  In this appendix we present the full proof of Theorem~\ref{thm.branching_theory_equiv}.

  Let us begin by recalling some conventions.
  The modality binds tightly on the left and loosely on the right;
  $\delta \wedge \gamma \tuple{\alpha} \varphi \wedge \psi$
  should be read as $\delta \wedge (\gamma \tuple{\alpha} (\varphi \wedge \psi))$.
  The set $\pmPHMLU$ is the set of PHMLU formulas and their negations
  and $\cpmPHMLU$ is the set of conjunctions of $\pmPHMLU$ formulas.

  Given a $\cpmPHMLU$ formula $\varphi$, we use $\varphi^+$ and $\varphi^-$
  to denote the decomposition of $\varphi$ into PHMLU formulas such that
  $\varphi$ is equivalent to $\varphi^+ \wedge \neg \varphi^-$.
  Concretely, $\varphi^+$ is the conjunction of the conjuncts of $\varphi$ that are PHMLU formulas,
  while $\varphi^-$ is the disjunction of those $\psi$ for which $\neg\psi$ is a conjunct of $\varphi$.

  Given formulas $\delta \in \PHMLU$ and $\varphi \in \cpmPHMLU$,
  we write $\delta \tuple{\alpha} \varphi$ to denote the PHMLU formula $\delta \tuple{\alpha} \varphi^+ \wedge \neg \varphi^-$.

  \begin{defi}
    We say that a formula $\varphi$ \emph{entails} a formula $\psi$ if
    in every state $s$ of every LTS,
    if $s \vDash \varphi$ then $s \vDash \psi$.
    We say $\varphi$ and $\psi$ are \emph{equivalent} if each entails the other.
  \end{defi}

  To aid in our proof, we define two functions
  \begin{align*}
    \Gamma_\alpha &: \HMLU^+ \times \HMLU \to \HMLU\\
    \Gamma^P_\alpha &: \left(\cpmPHMLU\right)^+ \times \cpmPHMLU \to \cpmPHMLU
  \end{align*}
  where $X^+$ denotes the set of finite non-empty sequences with elements from $X$.
  We define both functions recursively:
  \begin{align*}
    \Gamma_\alpha(\delta; \psi) &:= \delta \tuple{\alpha} \psi\\
    \Gamma_\alpha(\delta_1, \delta_2, \ldots, \delta_n; \psi) &:=
      \delta_1 \tuple{\tau} \Gamma_\alpha(\delta_2, \ldots, \delta_n; \psi)\\
    \Gamma^P_\alpha(\delta; \psi) &:= \neg\delta^- \wedge \delta^+ \tuple{\alpha} \psi\\
    \Gamma^P_\alpha(\delta_1, \delta_2, \ldots, \delta_n; \psi) &:=
       \neg\delta_1^- \wedge \delta_1^+ \tuple{\tau} \Gamma^P_\alpha(\delta_2, \ldots, \delta_n; \psi).
  \end{align*}

  We begin by proving a number of lemmas.

  \begin{lem}\label{lem.gamma_weakening}
    Given $\vec{\delta}, \vec{\theta} \in \HMLU^+$ and $\psi \in \HMLU$,
    with $\vec{\delta}$ and $\vec{\theta}$ of equal length and each $\delta_i$ entailing $\theta_i$,
    $\Gamma_\alpha(\vec{\delta}; \psi)$ entails $\Gamma_\alpha(\vec{\theta}; \psi)$.
  \end{lem}

  \begin{proof}
    For every state $s$, the path witnessing $s \vDash \Gamma_\alpha(\vec{\delta}; \psi)$ also witnesses 
    $s \vDash \Gamma_\alpha(\vec{\theta}; \psi)$.
  \end{proof}

  \begin{lem}\label{lem.gamma_one_step_equiv}
    Given $\delta, \psi \in \cpmPHMLU$, the formulas $\delta \tuple{\alpha} \psi$
    and $\neg\delta^- \wedge \delta^+ \tuple{\alpha} \psi$ are equivalent.
  \end{lem}

  \begin{proof}
    Note that this does not follow directly from $\delta$ and $\delta^+ \wedge \neg \delta^-$
    being equivalent: by our convention on parentheses, the formula 
    $\neg\delta^- \wedge \delta^+ \tuple{\alpha} \psi$ should be read as $\neg\delta^- \wedge \left(\delta^+ \tuple{\alpha} \psi\right)$.

    Let $s$ be an arbitrary state.
    If $s \vDash \delta \tuple{\alpha} \psi$, then in particular $s \vDash \delta$ and hence $s \vDash \neg\delta^-$.
    Moreover, the path $\satcommoneventual{s}{\delta}$ witnessing $s \vDash \delta \tuple{\alpha} \psi$
    also gives us $s \vDash \delta^+ \tuple{\alpha} \psi$.

    For the other direction, suppose $s \vDash \neg\delta^- \wedge \left(\delta^+ \tuple{\alpha} \psi\right)$.
    There is some path $\satcommoneventual{s}{\delta^+}$ witnessing the second conjunct, with $s'' \vDash \psi$.
    By Lemma~\ref{lem.transfer}, since $s \vDash \neg\delta^-$ and $\neg\delta^-$ is negative,
    every state on the path $s \rtctau^{\delta^+} s'$ satisfies $\neg\delta^-$,
    and hence $s \rtctau^\delta s'$.
    Thus $s \vDash \delta \tuple{\tau} \psi$, concluding the proof.
  \end{proof}

  \begin{lem}\label{lem.gamma_equiv}
    Given a sequence $\vec{\delta} \in \left(\cpmPHMLU\right)^+$ and a $\psi \in \cpmPHMLU$,
    $\Gamma^P_\alpha(\vec{\delta}; \psi)$ is equivalent to $\Gamma_\alpha(\vec{\delta}; \psi)$.
  \end{lem}

  \begin{proof}
    The argument proceeds by an induction on the length of $\vec{\delta}$.
    In both the base case and the inductive case, after writing out the definitions,
    the result follows from Lemma~\ref{lem.gamma_one_step_equiv}.
  \end{proof}

  Given a set $I$, let $\NRS(I)$ be the set of non-empty non-repeating sequences
  with elements from $I$.
  Note that if $I$ is finite, then so is $\NRS(I)$.

  \begin{lem}\label{lem.delta_shortening}
    Let $\vec{\delta} = \delta_1, \ldots, \delta_n$ and $\psi$ be $\cpmPHMLU$ formulas
    and suppose $\delta_{l_1} = \delta_{l_2}$ for $l_1 < l_2$.
    Then $s \vDash  \Gamma^P_\alpha(\vec{\delta}; \psi)$ entails 
    $s \vDash \Gamma^P_\alpha(\delta_1, \ldots, \delta_{l_1}, \delta_{l_2+1}, \ldots, \delta_n; \psi)$.
  \end{lem}

  \begin{proof}
    For this proof, the following notation is convenient:
    $$s \leadsto^{\delta}_\alpha t \qquad := \qquad \exists r.\, \sateventual{\alpha}{s}{r}{t}{\delta}.$$
    Suppose $s \vDash \Gamma^P_\alpha(\vec{\delta}; \psi)$.
    Unfolding the definitions, this means that there is a path
    \[
      s = s_1 \leadsto^{\delta^+_1}_\tau s_2 \leadsto^{\delta^+_2}_\tau \ldots \leadsto^{\delta^+_n}_\alpha s_{n+1}
    \]
    with for all $1 \le k \le n$, $s_k \vDash \neg\delta^-_k$, and with $s_{n+1} \vDash \psi$.

    We have $s_{l_1} \leadsto^{\delta^+_{l_1}}_\tau s_{l_2+1}$:
    since there is some $r$ such that $\sateventual{\tau}{s_{l_2}}{r}{s_{l_2+1}}{\delta^+_{l_2}}$,
    and since $\delta^+_{l_2}$ is positive, all states on this path leading up to $r$ satisfy $\delta^+_{l_2}$,
    which by assumption is equal to $\delta^+_{l_1}$.

    There is thus a path
    \[
      s = s_1 \leadsto^{\delta^+_1}_\tau \ldots \leadsto^{\delta^+_{l_1-1}}
        s_{l_1} \leadsto^{\delta^+_{l_1}}
        s_{l_2+1} \leadsto^{\delta^+_{l_2+1}} \ldots \leadsto^{\delta^+_n}_\alpha
        s_{n+1},
    \]
    where for all $k$ between $1$ and $l_1$ and between $l_2+1$ and $n$, $s_k \vDash \neg\delta^-_k$,
    which is a witness of $s \vDash \Gamma^P_\alpha(\delta_1, \ldots, \delta_{l_1}, \delta_{l_2+1}, \ldots, \delta_n; \psi)$,
    as desired.
  \end{proof}

  \begin{lem}\label{lem.gamma_disjunction}
    Given a finite family $(\delta_i)_{i \in I}$ of $\cpmPHMLU$ formulas and a $\cpmPHMLU$ formula $\psi$,
    the formula $\varphi = \left(\bigvee_i \delta_i\right)\tuple{\alpha} \psi$ is equivalent to
    \[
      \bigvee_{\vec{i} \in \NRS(I)} \Gamma^P_\alpha(\delta_{i_1}, \ldots, \delta_{i_n}; \psi).
    \]
  \end{lem}
  
  \begin{proof}
    Let $s$ be a state.
    We show that $\varphi$ holds in $s$ precisely when some $\vec{i} \in \NRS(I)$
    exists such that $s \vDash \Gamma^P_\alpha(\delta_{i_1}, \ldots, \delta_{i_n}; \psi)$.
    Let $\check{\delta}$ be a shorthand for $\bigvee_i \delta_i$.

    For the left-to-right direction, suppose $s \vDash \varphi$, and hence there exist $s', s''$
    such that $\satcommoneventual{s}{\check{\delta}}$ and $s'' \vDash \psi$.
    In particular, there exists a sequence $s = s_1 \totau \ldots \totau s_n = s'$
    such that for each $k \in \set{1, \ldots, n}$, $s_k \vDash \check{\delta}$.
    It follows that for every $k$ there is some disjunct with index $i_k$ such that $s_k \vDash \delta_{i_k}$.
    We thus see that
    $s \vDash \Gamma_\alpha(\delta_{i_1}, \ldots, \delta_{i_n}; \psi)$
    and by Lemma~\ref{lem.gamma_equiv}, we have
    $s \vDash \Gamma^P_\alpha(\delta_{i_1}, \ldots, \delta_{i_n}; \psi)$.

    We can now repeatedly use Lemma~\ref{lem.delta_shortening} to remove any duplicate indices
    that occur in $\vec{i}$, giving us $\vec{i'} \in \NRS(I)$ such that
    $s \vDash \Gamma^P_\alpha(\delta_{i'_1}, \ldots, \delta_{i'_n}; \psi)$.

    Conversely, suppose that there exists a $\vec{i} \in \NRS(I)$ such that
    $s \vDash \Gamma^P_\alpha(\delta_{i_1}, \ldots, \delta_{i_n}; \psi)$,
    and thus, by Lemma~\ref{lem.gamma_equiv},
    $s \vDash \Gamma_\alpha(\delta_{i_1}, \ldots, \delta_{i_n}; \psi)$.
    By Lemma~\ref{lem.gamma_weakening}, $\Gamma_\alpha(\delta_{i_1}, \ldots, \delta_{i_n}; \psi)$
    entails $\Gamma_\alpha(\check{\delta}, \ldots, \check{\delta}; \psi)$,
    which is equivalent to $\Gamma_\alpha(\check{\delta}; \psi) = \varphi$.
    It follows that $s \vDash \varphi$,
    which is exactly the desired conclusion.
  \end{proof}

  \begin{thm}[Restatement of Theorem~\ref{thm.branching_theory_equiv}]
    For every HMLU formula $\varphi$ there exists an equivalent disjunction of $\cpmPHMLU$ formulas.
    Moreover, if $\varphi$ is positive then it is equivalent to a disjunction of PHMLU formulas,
    and hence to a PHMLU formula.
  \end{thm}

  \begin{proof}
    The proof proceeds by induction, with all cases except the modality being straightforward.

    Suppose $\varphi = \delta \tuple{\alpha} \psi$.
    By induction, $\delta$ and $\psi$ are equivalent to $\check{\Delta} = \bigvee_i \Delta_i$ and $\bigvee_j \Psi_j$
    respectively.
    We define a family $\Phi$ indexed by $\NRS(I) \times J$ as follows:
    \[
      \Phi_{\vec{i}, j} := \Gamma^P_\alpha(\Delta_{i_1}, \ldots, \Delta_{i_n}; \Psi_j).
    \]

    It remains to show that $\bigvee_{\vec{i}, j} \Phi_{\vec{i}, j}$ is equivalent to $\varphi$.
    The chain of reasoning for this is as follows:
    \begin{align*}
      \check{\Delta}\tuple{\alpha}\bigvee_j\Psi_j
      &\Leftrightarrow
        \bigvee_j \left(\check{\Delta}\tuple{\alpha}\Psi_j\right)\\
      &\Leftrightarrow
        \bigvee_j \bigvee_{\vec{i} \in \NRS(I)} \Gamma^P_\alpha(\Delta_{i_1}, \ldots, \Delta_{i_n}; \Psi_j)\\
      &\Leftrightarrow
        \bigvee_{\vec{i}, j} \Gamma^P_\alpha(\Delta_{i_1}, \ldots, \Delta_{i_n}; \Psi_j).
    \end{align*}

    As in the case of Theorem~\ref{thm.strong_theory_equiv},
    the disjunction distributes over the modality when on the right.
    The next step holds by Lemma~\ref{lem.gamma_disjunction}, and finally we merge the two disjunctions.

    To show the second claim of the theorem, note that if $\varphi$ is positive, then so is $\delta$,
    and hence each $\Delta_i$ is a PHMLU formula.
    We thus take $\Delta_i^+ = \Delta_i$ and $\Delta_i^- = \bot$.
    It follows that each $\Phi_{\vec{i}, j}$ is equivalent to a PHMLU formula, as desired.
  \end{proof}

 \section{Algorithm and Example}\label{app.example}
  We now describe the algorithm that computes a distinguishing PHMLU
  formula from a derivation of a directed apartness and the reverse
  algorithm that computes a derivation of $s\dApart t$ from a PHMLU
  formula $\varphi$ for which $s \vDash \varphi$ and $t\not\vDash
  \varphi$. This is the algorithm implicit in the proof of Theorem~\ref{thm.directed_branching_hm}
  that ties together Lemma~\ref{lem.dirbapt},
  describing derivations of $s\dApart t$, and Lemma~\ref{lem.positive_semantics}, describing $s\vDash \varphi$.

  \paragraph*{From a derivation of $s \dApart t$ to a PHMLU formula}
A derivation of $s \dApart t$ has as last rule
  \[
    \inferrule*[Right=$\textsc{in}_{db}$]{
      s \rtctau s' \to_\alpha s'' \\
      \forall t', t''.\, \commoneventual{t} \implies s' \dApart t' \vee s'' \dApartSym t''
    }{s \dApart t}
  \]
  We consider the (finite) set of hypotheses of the rule, which
  are all of the form (1) $s' \dApart t'$ or (2) $s'' \dApart t''$ or
  (3) $t'' \dApart s''$ (where $\commoneventual{t}$) For
  each hypothesis we have (recursively) a PHMLU formula that
  distinguishes the states and we collect them together: $\delta$ is
  the conjunction of the formulas that we obtain from hypotheses of
  the form (1), $\varphi_1$ is the conjunction of the formulas that we
  obtain from hypotheses of the form (2) and $\varphi_2$ is the
  disjunction of the formulas that we obtain from hypotheses of the
  form (3). Now take $\varphi := \delta \tuple{\alpha} (\varphi_1 \wedge \neg
  \varphi_2)$. Then $s\vDash \varphi$ and $t\nvDash \varphi$.

  \paragraph*{From a PHMLU formula distinguishing $s$ and $t$ to a derivation of $s \dApart t$}
Given the PHMLU formula $\varphi$ for which $s\vDash \varphi$ and
$t\not\vDash \varphi$, we compute, recursively, a derivation of
$s\dApart t$. The cases for $\varphi = \bot, \top$ cannot occur, and
the cases for $\varphi = \varphi_1 \wedge \varphi_2$ or $\varphi = \varphi_1
\vee \varphi_2$ are dealt with by an immediate recursive call. (For the
case of $\varphi = \varphi_1 \wedge \varphi_2$, we have $t\nvDash\varphi_1 \wedge
\varphi_2$, so $t\nvDash \varphi_1$ or $t\nvDash \varphi_2$ (possibly
both). Say we have $t\nvDash \varphi_1$. We also have $s\vDash \varphi_1$,
so we take $\varphi := \varphi_1$.)

For the case of $\varphi := \delta \tuple{\alpha} (\varphi_1 \wedge
\neg \varphi_2)$, from $s \vDash \varphi$ we find $s', s''$ with
$\commoneventual{s}$ and $s' \vDash \delta$, $s'' \vDash
\varphi_1$ and $s'' \nvDash \varphi_2$. From $t \nvDash \varphi$ we
find that for all $t', t''$ with $\commoneventual{t}$ we have
$t' \nvDash \delta$ or $t'' \nvDash \varphi_1$ or $t'' \vDash
\varphi_2$. We recursively have, for all $t', t''$ with $\commoneventual{t}$,
a derivation of $s' \dApart t'$ or of $s'' \dApart
t''$ or of $t'' \dApart s''$, so by rule ($\textsc{in}_{db}$) we have
a derivation of $s \dApart t$.
  
  We now give another slightly longer example of how a distinguishing
  formula can be computed from an apartness proof and vice versa. We
  give an example of the other direction then the one that was
  shown in Example~\ref{exa.LTSbranapt}, we show how a derivation of
  $t_0\dApart s_0$ is computed from the distinguishing formula.
  
    \begin{exa}\label{exa.fromGeuvers2}
  For the LTS given below we have $t_0\dApart s_0$.
  We give a distinguishing formula and from that produce a derivation of $t_0\dApart s_0$,
  following the proof of Theorem~\ref{thm.directed_branching_hm}.
  
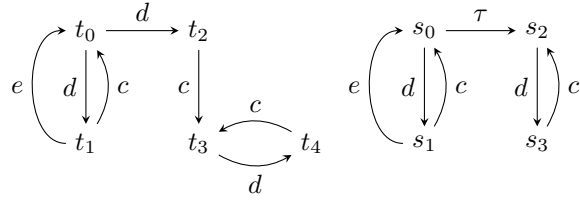
  \begin{center}
  {\sizeChoice
    \begin{tabular}[t]{cc}
      \begin{tikzpicture}[>=stealth,node distance=1.5cm,auto]
          \node  (q0)                        {$t_0$};
          \node  (q1) [below of = q0]        {$t_1$};
          \node  (q2) [right of = q0]        {$t_2$};
          \node  (q3) [below of = q2]        {$t_3$};
          \node  (q4) [right of = q3]        {$t_4$};
          \node  (p1) [right of = q4]        {$s_1$};
          \node  (p0) [above of = p1]        {$s_0$};
          \node  (p2) [right of = p0]        {$s_2$};
          \node  (p3) [below of = p2]        {$s_3$};
          \path[->]
          (q0) edge                   node[swap] {$d$}    (q1)
              edge                   node {$d$}       (q2)
          (q1) edge [bend right]      node[swap] {$c$}    (q0)
              edge [bend left=90]    node {$e$}          (q0)
          (q2) edge                   node[swap] {$c$}    (q3)
          (q3) edge [bend right]      node[swap] {$d$}    (q4)
          (q4) edge [bend right]      node[swap] {$c$}    (q3)
          (p0) edge                   node[swap] {$d$}    (p1)
              edge                   node {$\tau$}       (p2)
          (p1) edge [bend right]     node[swap] {$c$}    (p0)
              edge [bend left=90]     node {$e$}          (p0)
          (p2) edge      node[swap] {$d$}    (p3)
          (p3) edge [bend right]      node[swap] {$c$}          (p2);
      \end{tikzpicture}
    \end{tabular}
  }
  \end{center}
  
  A distinguishing PHMLU-formula for $t_0$ and $s_0$ is
$$\varphi := (\tuple{d}\,\tuple{e}\,\top) \;\tuple{d}\; (\neg\tuple{e}\,\top).$$
From $\varphi$ we can compute a derivation of $t_0\dApart s_0$, following the algorithm that we have just described.
For purposes of space, we single out a sub-derivation $\Sigma$ of $t_0\dApart s_2$.
We indicate above the $\forall s', s'' \ldots$ all the (finitely many) possible transitions
that we need to prove a property for: two in the case of $\eventual{d}{s_0}{s'}{s''}$, and  one in the case of $\eventual{d}{s_2}{s'}{s''}$.
Similarly above the $\forall r', r'' \ldots$, but there are no possible transitions, so there is nothing to prove, therefore $\checkmark$.

Note that, to compute the derivation from the formula, we need to know
the actual reason why $t_0 \vDash \varphi$ and $s_0 \nvDash \varphi$,
because we construct the derivation from that information. For $t_0
\vDash \varphi$ we consider $t_0 \maybeto{d} t_2$, using which we create a
derivation with two subderivations: one basically with conclusion $s_1
\dApart t_2$ (corresponding with formula $\neg\tuple{e}\,\top$) and
the other, $\Sigma$, with conclusion $t_0 \dApart s_2$ (corresponding
with formula $\tuple{d}\,\tuple{e}\,\top$).

{\sizeChoice
    \[
      \inferrule*{
        t_0 \to_d t_2 \quad 
        \inferrule*{
          \inferrule*{
            \inferrule*{
              \inferrule*{
                s_1 \to_e s_0 \quad
                \inferrule*{\checkmark}
                {\forall r',r''.\, \eventual{e}{t_2}{r'}{r''} \implies \ldots}
              } {s_1 \dApart t_2}
            } {t_2 \dApartSym s_1}
          } {t_0 \dApart s_0 \vee t_2 \dApartSym s_1}
          \\
          \inferrule*{
            \Sigma
          } {t_0 \dApart s_2 \vee t_2 \dApartSym s_3}
        } {\forall s',s''.\, \eventual{d}{s_0}{s'}{s''}\implies t_0 \dApart s'\vee t_2 \dApartSym s''}
      } {t_0\dApart s_0}
    \]
$\Sigma:=$
    \[
      \inferrule*{
        t_0 \to_d t_1 \\
        \inferrule*{
          \inferrule*{
            \inferrule*{
              \inferrule*{
                t_1 \to_e t_0 \quad
                \inferrule*{\checkmark}
                {\forall r',r''.\, \eventual{e}{s_3}{r'}{r''} \implies \ldots}
              } {t_1 \dApart s_3}
            } {t_1 \dApartSym s_3}
          } {t_0 \dApart s_2 \vee t_1 \dApartSym s_3}
        } {\forall s',s''.\, \eventual{d}{s_2}{s'}{s''} \implies t_0 \dApart s'\vee t_1 \dApartSym s''}
      } {t_0 \dApart s_2}
    \]

  }

\end{exa}
\end{document}